\documentstyle[12pt,aaspp]{article}

\begin{document}

\title{A THREE-DIMENSIONAL HYDRODYNAMICS CODE FOR MODELING
SOURCES OF GRAVITATIONAL RADIATION}
\author{Scott C. Smith,{\altaffilmark{1,2}} Joan M. Centrella,
{\altaffilmark{1}} and Sean P. Clancy{\altaffilmark{3}}}

\altaffiltext{1}{Dept. of Physics and Atmospheric Science
Drexel University, Philadelphia, PA, 19104}
\altaffiltext{2}{Dept. of Physics, Muhlenberg College, Allentown,
PA 18104}
\altaffiltext{3}{X-3, MS-F663, Los Alamos National Laboratory,
Los Alamos, NM  87545}

\begin{abstract}
We have developed a 3-D Eulerian hydrodynamics code to model sources
of gravitational radiation.  The code is written in cylindrical
coordinates $(\varpi,z,\varphi)$ and has moving grids in the $\varpi$
and $z$-directions.  We use Newtonian gravity and calculate the
gravitational radiation in the quadrupole
approximation.  This code has been tested on a variety of problems to
verify its accuracy and stability, and the results of these tests are
reported here.
\end{abstract}


\section{INTRODUCTION}

Many of the most interesting astrophysical systems can be described by
the equations of hydrodynamics coupled to gravity.  Among these are
sources of gravitational waves such as inspiralling neutron star
binaries and collapsing compact stars undergoing global rotational
instabilities (Thorne 1987, 1992; Schutz 1989).  With the prospect of
several gravitational wave detectors becoming operational within a
decade (see Abramovici, et al.  1992 and references therein), the
detailed modeling of such sources has a high priority.  Since these
are time-dependent, nonlinear, and fully 3-D systems, calculating the
gravitational radiation they produce requires numerical simulations.

We have developed a 3-D Eulerian hydrodynamics code to model such
sources.  The code is written in cylindrical coordinates with a moving
grid.  It uses Newtonian gravity and calculates the gravitational
radiation produced in the quadrupole limit.  The code is structured so
that post-Newtonian effects, and even the full field equations of
general relativity, can be coupled to the basic hydrodynamics routines
in the future.  In this paper we describe this code and the tests we
have performed to insure its accuracy.  Section~\ref{hydro-eqs}
presents the basic hydrodynamical equations on which the code is built
and \S~\ref{sec-grid-cen} discusses the numerical grids and the
centering of the basic variables on those grids.  The numerical
implementation of the hydrodynamical equations is presented in detail
in \S~\ref{sec-FD}, including a discussion of the finite differencing
techniques used.  The method used to solve Poisson's equation is
discussed in \S~\ref{sec-pois}.  Section~\ref{extract} considers the
extraction of gravitational radiation using various quadrupole
formulae.  The test-bed problems we ran on the code and the results we
obtained are presented in \S~\ref{tests}.  Finally, \S~\ref{summary}
summarizes our work.

\section{HYDRODYNAMICAL EQUATIONS}\label{hydro-eqs}

We begin with the basic equations of hydrodynamics (Landau \& Lifshitz
1959; Bowers \& Wilson 1991).  Conservation of mass gives the mass
continuity equation
\begin{equation}
\frac{d\rho}{dt} \:+\: \rho\,\nabla\cdot\vec{v}\;=\;0,
\label{mass-cont-L}
\end{equation}
where $d/dt$ is the total or Lagrangian time derivative.  For a moving
coordinate system
\begin{equation}
\frac{d}{dt}=\frac{\partial}{\partial t} + \vec{v}_r\cdot\nabla,
\end{equation}
where $\vec v = \vec v_r + \vec v_g$ is the velocity of the fluid,
$\vec{v}_r$ is the velocity of the fluid relative to the grid, and
$\vec{v}_g$ is the velocity of the grid.  Substituting this
into~(\ref{mass-cont-L}) gives
\begin{equation}
\frac{\partial\rho}{\partial
t}\:+\:\nabla\cdot(\rho\vec{v}_r)\:+\:\rho\nabla\cdot\vec{v}_g\;=\;0,
\label{mass-cont}
\end{equation}
which is the mass continuity equation with a moving grid.  The
remaining equations will all be given in a form which shows the moving
grid terms explicitly.  The momentum conservation equation, or Euler's
equation, is
\begin{equation}
\frac{\partial(\rho\vec{v})}{\partial t}\:+\:\rho\vec{v}\nabla\cdot\vec{v}_g
\;=\;-\nabla\cdot(\rho\vec{v}\vec{v}_r)\:-\:\nabla P\:+\:\vec{f},
\end{equation}
or, in component form,
\begin{equation}
\frac{\partial(\rho v^k)}{\partial t}\:+\:\rho v^k\nabla\cdot\vec{v}_g
\;=\;-\nabla\cdot(\rho v^k\vec{v}_r)\:-\:\nabla_k P\:+\:f^k ,
\label{euler}
\end{equation}
where $\vec f$ is the body force density.  The body force can arise
from many sources, including self-gravity of the fluid, electric and
magnetic fields generated within the fluid, and externally applied
fields.  In the current version of the code, we are only concerned
with the body force density due to the self-gravity of the fluid.
This is given by
\begin{equation}
\vec{f}\;=\;-\rho\nabla\Phi,
\label{gforce}
\end{equation}
where the gravitational potential $\Phi$ satisfies Poisson's equation
\begin{equation}
\nabla^2\Phi\;=\;4\pi G\rho.
\label{poisson}
\end{equation}
We follow Bowers \& Wilson (1991) and use an internal energy equation
\begin{equation}
\frac{\partial (\rho\varepsilon)}{\partial t}\;=\;
-\nabla\cdot(\rho\varepsilon\vec{v}_r)\:-\:\rho\varepsilon\nabla\cdot\vec{v}_g
\:-\:P\nabla\cdot\vec{v} ,
\label{energy}
\end{equation}
where $\epsilon$ is the specific internal energy of the fluid.  To
complete this set we need the equation of state (EOS) of the fluid,
$P\;=\;P(\rho,\varepsilon)$.  In this paper we use the EOS for a
perfect fluid,
\begin{equation}
P\;=\;(\gamma-1)\rho\varepsilon .
\label{ideal-gas}
\end{equation}
More sophisticated equations of state, in the form of analytic
expressions or tabulated values, can be inserted into the code in the
future.

We use a cylindrical coordinate system in our code since it gives an
efficient means of modeling rotating systems; in addition stellar
collisions and inspiralling binary stars can easily be set up.  To
formulate the hydrodynamical equations in cylindrical coordinates
$(\varpi,z,\varphi)$ we write $\vec{v}=(v,u,w)$, with similar
expressions for $\vec{v}_g$ and $\vec{v}_r$ (note that $w$ is a {\it
linear} velocity so that $w=\varpi\Omega$, where $\Omega$ is the
angular velocity).  With this, and the definition of the gradient in
cylindrical coordinates, the hydrodynamical equations take the
following forms (Clancy 1989; Smith 1993).  The mass continuity
equation~(\ref{mass-cont}) becomes
\begin{equation}
\frac{\partial\rho}{\partial t}=
-\frac{1}{\varpi}\frac{\partial(\varpi\rho v_r)}{\partial \varpi}
-\frac{\partial(\rho u_r)}{\partial z}
-\frac{1}{\varpi}\frac{\partial(\rho w_r)}{\partial\varphi} -\rho
\left(\frac{1}{\varpi}\frac{\partial(\varpi v_g)}{\partial\varpi}
+\frac{\partial u_g}{\partial z} +\frac{1}{\varpi}\frac{\partial
w_g}{\partial\varphi}\right)
\label{mass-cyl}
\end{equation}
and the internal energy equation~(\ref{energy}) is\pagebreak[1]
{\samepage
\begin{eqnarray}
\frac{\partial(\rho\varepsilon)}{\partial t} & = &
-\frac{1}{\varpi}\frac{\partial(\varpi\rho\varepsilon v_r)}{\partial
\varpi}
\:-\:\frac{\partial(\rho\varepsilon u_r)}{\partial z}
\:-\:\frac{1}{\varpi}\frac{\partial(\rho\varepsilon w_r)}{\partial\varphi}
\nonumber \\ & & \;\;\;
\:-\:\rho\varepsilon\left(\frac{1}{\varpi}\frac{\partial(\varpi
v_g)}{\partial\varpi}
\,+\,\frac{\partial u_g}{\partial z}
\,+\,\frac{1}{\varpi}\frac{\partial w_g}{\partial\varphi}\right) \nonumber \\ &
& \;\;\;
\:-\:P\left(\frac{1}{\varpi}\frac{\partial(\varpi v)}{\partial\varpi}
\,+\,\frac{\partial u}{\partial z}
\,+\,\frac{1}{\varpi}\frac{\partial w}{\partial\varphi}\right).
\label{ener-cyl}
\end{eqnarray}}
Euler's equation (\ref{euler}) in component form
becomes\pagebreak[1]{\samepage\begin{eqnarray}
\frac{\partial(\rho v)}{\partial t} & = &
-\frac{1}{\varpi}\frac{\partial(\varpi\rho v v_r)}{\partial \varpi}
\:-\:\frac{\partial(\rho v u_r)}{\partial z}
\:-\:\frac{1}{\varpi}\frac{\partial(\rho v w_r)}{\partial\varphi}
\nonumber \\ & & \;\;
-\:\rho v\left(\frac{1}{\varpi}\frac{\partial(\varpi
v_g)}{\partial\varpi} +\frac{\partial u_g}{\partial z}
+\frac{1}{\varpi}\frac{\partial w_g}{\partial\varphi}\right)
\:-\:\frac{\partial
P}{\partial\varpi}\:-\:\rho\frac{\partial\Phi}{\partial\varpi},
\label{euler-r}
\end{eqnarray}}\pagebreak[1]
{\samepage\begin{eqnarray}
\frac{\partial(\rho u)}{\partial t} & = &
-\frac{1}{\varpi}\frac{\partial(\varpi\rho u v_r)}{\partial \varpi}
\:-\:\frac{\partial(\rho u u_r)}{\partial z}
\:-\:\frac{1}{\varpi}\frac{\partial(\rho u w_r)}{\partial\varphi}
\nonumber \\ & & \;\;
-\:\rho u\left(\frac{1}{\varpi}\frac{\partial(\varpi
v_g)}{\partial\varpi} +\frac{\partial u_g}{\partial z}
+\frac{1}{\varpi}\frac{\partial w_g}{\partial\varphi}\right)
\:-\:\frac{\partial P}{\partial z}\:-\:\rho\frac{\partial\Phi}{\partial z},
\label{euler-z}
\end{eqnarray}}\pagebreak[1]
and {\samepage\begin{eqnarray}
\frac{\partial(\rho w)}{\partial t} & = &
-\frac{1}{\varpi}\frac{\partial(\varpi\rho w v_r)}{\partial \varpi}
\:-\:\frac{\partial(\rho w u_r)}{\partial z}
\:-\:\frac{1}{\varpi}\frac{\partial(\rho w w_r)}{\partial\varphi}
\nonumber \\ & & \;\;
-\rho w\left(\frac{1}{\varpi}\frac{\partial(\varpi
v_g)}{\partial\varpi} +\frac{\partial u_g}{\partial z}
+\frac{1}{\varpi}\frac{\partial w_g}{\partial\varphi}\right)
-\frac{1}{\varpi}\frac{\partial P}{\partial\varphi}
-\frac{\rho}{\varpi}\frac{\partial\Phi}{\partial\varphi}.
\label{euler-phi}
\end{eqnarray}}\pagebreak[1]
Poisson's equation (\ref{poisson}) for the potential $\Phi$ in
cylindrical coordinates is
\begin{equation}
\frac{1}{\varpi}\frac{\partial}{\partial\varpi}\left(\varpi\frac{\partial\Phi}{\partial\varpi}\right)
\:+\:\frac{\partial^2\Phi}{\partial z^2}\:+\:
\frac{1}{\varpi^2}\frac{\partial^2\Phi}{\partial\varphi^2}\;=\;4\pi G\rho.
\label{poisson-cyl}
\end{equation}
 The EOS is independent of coordinates, and remains unchanged.  Thus,
(\ref{mass-cyl})--(\ref{poisson-cyl}), along with (\ref{ideal-gas}),
are the equations that will be implemented in the hydrodynamics code.

\section{NUMERICAL GRIDS AND CENTERING OF VARIABLES}
\label{sec-grid-cen}

The next step is to define the numerical grids in space and time on
which the hydrodynamical equations will be discretized.  We use
staggered spatial and temporal grids, and place the physical variables
in such a way as to simplify calculations (Bowers \& Wilson 1991).

For each of the cylindrical coordinates $(\varpi,z,\varphi)$ we define
two grids: a ``zone centered'' grid, which has nodes at the centers of
grid zones, and a ``face centered'' grid, which has nodes on the faces
or boundaries separating grid zones.  This is illustrated for the
axial or z-grid in Figure~\ref{z-gridfig}.  In our notation, grid zone
boundaries are denoted by integral indices and zone centers by
half-integral indices. Thus, for an array of zone boundaries
$\{z_k\}$, the center of the $k$th axial zone is given by
\begin{displaymath}
z_{k+\frac{1}{2}}\;=\;\frac{1}{2}\left(z_k + z_{k+1} \right),
\end{displaymath}
and the grid spacings are $$
\Delta z_{k+\frac{1}{2}}\;=\;\frac{1}{2}\left(z_{k+1} - z_{k} \right)
$$ and $$
\Delta z_{k}\;=\;\frac{1}{2}\left(z_{k+\frac{1}{2}} - z_{k-\frac{1}{2}}
\right).
$$ The radial and angular grids are handled in a similar manner.  For
example, the $j$th radial zone is bounded by $\varpi_j$ and
$\varpi_{j+1}$, and has its center at
\begin{displaymath}
\varpi_{j+\frac{1}{2}} \;=\;\frac{1}{2}\left(\varpi_j\:+\:\varpi_{j+1}\right).
\end{displaymath}
The evolution of the fluid is carried out on grids that are staggered
in time.  For reasons that will become evident later, the zone
centered quantities are defined at integral time steps, while face
centered variables are defined at half-integral time steps.

Throughout this paper, we use the subscript indices $j,k,l$ and the
superscript $n$ to refer to the grids in $\varpi ,z,\varphi$, and $t$,
respectively.  Thus, a zone centered quantity $S$ is written
\begin{displaymath}
S^n_{j+\frac{1}{2},k+\frac{1}{2},l+\frac{1}{2}} \;\equiv\;
S(\varpi_{j+\frac{1}{2}},z_{k+\frac{1}{2}},
\varphi_{l+\frac{1}{2}},t^n),
\end{displaymath}
and a face centered quantity $V$ is
\begin{displaymath}
V^{n+\frac{1}{2}}_{j,k+\frac{1}{2},l+\frac{1}{2}} \;\equiv\;
V(\varpi_{j},z_{k+\frac{1}{2}},
\varphi_{l+\frac{1}{2}},t^{n+\frac{1}{2}}).
\end{displaymath}
Any variable that has three half-integral spatial indices is defined
at the center of a 3-D grid cell.  Variables having two half-integral
indices and one integral index are defined in the center of a zone
face (the surface separating two neighboring zones).  While it is
occasionally necessary to evaluate a quantity at a zone edge (where
four zones meet, specified by one half-integral and two integral
indices), physical quantities are never evaluated at zone corners (all
integral indices).

In order to reduce memory requirements we have chosen to let our grid
cover only the ``top'' of the coordinate system $z\geq 0$, which
restricts us to treating problems for which $z=0$ is a plane of
reflection symmetry.  This is not currently a serious limitation, as
there is a large class of interesting problems possessing such a plane
of symmetry.  Furthermore, this restriction can be removed in the
future with a small number of changes.

In our code scalar quantities, such as mass, density, and
gravitational potential, are defined at the zone centers and at
integral time steps.  Vector quantities, such as velocities, are
defined on the faces between zones and at half-integral time steps.
This is the most natural choice for locating the variables, since the
velocities are updated using the gradients of scalar quantities
(pressure and gravitational potential), which are naturally defined on
the zone faces, and the velocities are used to transport scalar
quantities between zones.  The velocity components are face-centered
in the coordinate along which they are directed, as illustrated in
Figure~\ref{gridfig}.

The physical variables thus map onto the spatial and temporal
numerical grids as follows:
\begin{tabbing}
{}~~~~~~~\= density:~~~~~~~~~~~~~~~~~~~~~~~~~\=
$\rho(\varpi,z,\varphi,t)$~~~ \= $\rightarrow$~~\=
$\rho^{n}_{j+\frac{1}{2},k+\frac{1}{2},l+\frac{1}{2}}$\\
\>specific internal energy: \> $\varepsilon(\varpi,z,
\varphi,t)$ \> $\rightarrow$ \> $\varepsilon^{n}_{j+
\frac{1}{2},k+\frac{1}{2},l
+\frac{1}{2}}$\\
\> pressure: \> $p(\varpi,z,\varphi,t)$ \> $\rightarrow$ \>
$p^{n}_{j+\frac{1}{2},k+\frac{1}{2},l+\frac{1}{2}}$\\
\> gravitational potential: \> $\Phi(\varpi,z,\varphi,t)$
 \> $\rightarrow$ \> $\Phi^{n}_{j+\frac{1}{2},k+
\frac{1}{2},l+\frac{1}{2}}$\\
\> fluid velocity: \> $v_{\varpi}(\varpi,z,\varphi,t)$
\> $\rightarrow$ \> $v^{n+\frac{1}{2}}_{j,k+
\frac{1}{2},l+\frac{1}{2}}$\\
\> \> $v_{z}(\varpi,z,\varphi,t)$ \> $\rightarrow$ \>
$u^{n+\frac{1}{2}}_{j+\frac{1}{2},k,l+\frac{1}{2}}$\\
\> \> $v_{\varphi}(\varpi,z,\varphi,t)$ \> $\rightarrow$ \>
$w^{n+\frac{1}{2}}_{j+\frac{1}{2},k+\frac{1}{2},l}$\\
\> grid velocity: \> ${v_g}_{\varpi}(\varpi,t)$ \>
$\rightarrow$ \> ${v_g}^{n+\frac{1}{2}}_{j}$,
${v_g}^{n+\frac{1}{2}}_{j+\frac{1}{2}}$\\
\> \> ${v_g}_{z}(z,t)$ \> $\rightarrow$ \>
 ${u_g}^{n+\frac{1}{2}}_{k}$, ${u_g}^{n+
\frac{1}{2}}_{k+\frac{1}{2}}$ \\
\> \> ${v_g}_{\varphi}(\varphi,t)$ \> $\rightarrow$ \>
${w_g}_{l}^{n+\frac{1}{2}}$,
${w_g}_{l+\frac{1}{2}}^{n+\frac{1}{2}}$\\
\end{tabbing}
Note that each component of the grid velocity maps into two 1-D
arrays, one for the zone centered grid and one for the face centered
grid in that coordinate.

It is sometimes necessary to evaluate a zone centered quantity on a
face centered grid; we accomplish this by using a volume average
(Clancy 1989).  Consider the case of averaging a zone centered
quantity $q_{j+\frac{1}{2},k+\frac{1}{2},l+\frac{1}{2}}$ (temporal
index suppressed) in the axial direction to obtain the value
$q_{j+\frac{1}{2},k,l+\frac{1}{2}}$.  Zone centered quantities are
defined to be constant throughout the zone, as indicated in
Figure~\ref{fig-average}.  The volume averaged value of $q$ in the
$z$-direction is given by
\begin{equation}
q_{j+\frac{1}{2},k,l+\frac{1}{2}} = \frac{1}{2} \;
\frac{q_{j+\frac{1}{2},k-\frac{1}{2},l+\frac{1}{2}}\Delta
V_{j+\frac{1}{2},k-\frac{1}{2},l+\frac{1}{2}} +
q_{j+\frac{1}{2},k+\frac{1}{2},l+\frac{1}{2}}\Delta
V_{j+\frac{1}{2},k+\frac{1}{2},l+\frac{1}{2}}} {\Delta
V_{j+\frac{1}{2},k,l+\frac{1}{2}}},
\label{vol-avg}
\end{equation}
which reduces to a simple average if all volume elements are equal.
Since the volume element is
\begin{equation}
\Delta V_{j+\frac{1}{2},k+\frac{1}{2},l+\frac{1}{2}} =
\varpi_{j+\frac{1}{2}}\Delta\varpi_{j+\frac{1}{2}}
\,\Delta z_{k+\frac{1}{2}} \,\Delta\varphi_{l+\frac{1}{2}},
\label{dvol}\end{equation}
equation~(\ref{vol-avg}) reduces to
\begin{equation}
q_{j+\frac{1}{2},k,l+\frac{1}{2}} = \frac{1}{2} \;
\frac{q_{j+\frac{1}{2},k-\frac{1}{2},l+\frac{1}{2}}\Delta z_{k-\frac{1}{2}} +
q_{j+\frac{1}{2},k+\frac{1}{2},l+\frac{1}{2}}\Delta z_{k+\frac{1}{2}}}
{\Delta z_{k}}.
\label{z-vol-avg}
\end{equation}
Similarly, volume averaging in the angular direction gives
\begin{equation}
q_{j+\frac{1}{2},k+\frac{1}{2},l} = \frac{1}{2} \;
\frac{q_{j+\frac{1}{2},k+\frac{1}{2},l-\frac{1}{2}}\Delta\varphi_{l-\frac{1}{2}} +
q_{j+\frac{1}{2},k+\frac{1}{2},l+\frac{1}{2}}\Delta\varphi_{l+\frac{1}{2}}}
{\Delta\varphi_{l}}.
\label{t-vol-avg}
\end{equation}

A problem arises if this simple prescription is followed for volume
averaging in the $\varpi$-direction.  The radial averaged analog of
(\ref{vol-avg}) is
\begin{equation}
q_{j,k+\frac{1}{2},l+\frac{1}{2}} = \frac{1}{2} \;
\frac{q_{j-\frac{1}{2},k+\frac{1}{2},l+\frac{1}{2}}(\varpi\Delta\varpi)_{j-\frac{1}{2}} +
q_{j+\frac{1}{2},k+\frac{1}{2},l+\frac{1}{2}}(\varpi\Delta\varpi)_{j+\frac{1}{2}}}
{\varpi_j\Delta\varpi_{j}};
\label{r-vol-avg}
\end{equation}
however, this expression will introduce inaccuracy arising from the
curvature of the coordinate system.  For example, if
$q_{j+\frac{1}{2}} = q_{j-\frac{1}{2}}$ (angular and axial subscripts
suppressed), then the face centered quantity $q_j$ should have this
same value.  If (\ref{r-vol-avg}) is applied, however, we see that
$q_j$ differs from the values on either side by a factor $$
\frac{q_j}{q_{j+\frac{1}{2}}} = \frac{1}{2} \;
\frac{(\varpi\Delta\varpi)_{j-\frac{1}{2}} +
(\varpi\Delta\varpi)_{j+\frac{1}{2}}}{\varpi_j\Delta\varpi_{j}} = 1 +
\frac{1}{2}\frac{\Delta\varpi_{j+\frac{1}{2}}-\Delta\varpi_{j-\frac{1}{2}}}{\varpi_j}.
$$ The numerical error present in this term is particularly pronounced
near the axis coordinate singularity.  However, it can easily be
avoided by replacing the radius $\varpi_j$ in the denominator of
(\ref{r-vol-avg}) with a curvature corrected radius
$\widetilde{\varpi}_j$ given by
\begin{equation}
\widetilde{\varpi}_j \equiv
\frac{1}{2}\frac{(\varpi\Delta\varpi)_{j-\frac{1}{2}} +
(\varpi\Delta\varpi)_{j+\frac{1}{2}}}{\Delta\varpi_{j}},
\label{rtilde}
\end{equation}
to yield
\begin{equation}
q_{j,k+\frac{1}{2},l+\frac{1}{2}} = \frac{1}{2} \;
\frac{q_{j-\frac{1}{2},k+\frac{1}{2},l+\frac{1}{2}}(\varpi\Delta\varpi)_{j-\frac{1}{2}} +
q_{j+\frac{1}{2},k+\frac{1}{2},l+\frac{1}{2}}(\varpi\Delta\varpi)_{j+\frac{1}{2}}}
{\widetilde{\varpi}_j\Delta\varpi_{j}}.
\label{rt-vol-avg}
\end{equation}
Using equation~(\ref{rt-vol-avg}) provides a more accurate volume
average in the radial direction, and helps to avoid singular behavior
on the coordinate axis.

It is also necessary to evaluate face centered quantities at zone
centers, and this is accomplished using expressions analogous to
(\ref{z-vol-avg}), (\ref{t-vol-avg}), (\ref{rtilde}), and
(\ref{rt-vol-avg}).  These are given by
\begin{equation}
q_{j+\frac{1}{2},k+\frac{1}{2},l+\frac{1}{2}} = \frac{1}{2} \;
\frac{q_{j+\frac{1}{2},k,l+\frac{1}{2}}\Delta z_{k} +
q_{j+\frac{1}{2},k+1,l+\frac{1}{2}}\Delta z_{k+1}} {\Delta
z_{k+\frac{1}{2}}},
\end{equation}
\begin{equation}
q_{j+\frac{1}{2},k+\frac{1}{2},l+\frac{1}{2}} = \frac{1}{2} \;
\frac{q_{j+\frac{1}{2},k+\frac{1}{2},l}\Delta\varphi_{l} +
q_{j+\frac{1}{2},k+\frac{1}{2},l+1}\Delta\varphi_{l+1}}
{\Delta\varphi_{l+\frac{1}{2}}},
\end{equation}
and
\begin{equation}
q_{j+\frac{1}{2},k+\frac{1}{2},l+\frac{1}{2}} = \frac{1}{2} \;
\frac{q_{j,k+\frac{1}{2},l+\frac{1}{2}}(\varpi\Delta\varpi)_{j} +
q_{j+1,k+\frac{1}{2},l+\frac{1}{2}}(\varpi\Delta\varpi)_{j+1}}
{\widetilde{\varpi}_{j+\frac{1}{2}}\Delta\varpi_{j+\frac{1}{2}}},
\end{equation}
where
\begin{equation}
\widetilde{\varpi}_{j+\frac{1}{2}} \equiv \frac{1}{2} \;
\frac{(\varpi\Delta\varpi)_{j} +
(\varpi\Delta\varpi)_{j+1}}{\Delta\varpi_{j+\frac{1}{2}}}.
\end{equation}

\section{FINITE DIFFERENCED HYDRO EQUATIONS}
\label{sec-FD}

In this section we obtain the operator split, finite difference
versions of the Eulerian hydrodynamical equations (\ref{mass-cyl}) -
(\ref{poisson-cyl}).  We assume that the values of the scalar
variables are known at time step $n$ and that the velocities are known
at time step $n\!-\!\frac{1}{2}$.

\subsection{Code Architecture}

When constructing a general purpose hydrodynamics code it is important
to maintain flexibility and allow for future modifications by using
structured programming techniques.  Our code is divided into a large
number of separate subroutines, each of which is dedicated to a
specific task and may be replaced or modified as needed for particular
astrophysical problems (Bowers \& Wilson 1991).  This modular
structure allows the program to be easily adapted to study new
situations, adding or removing different processes as desired, and
makes it possible to change a subroutine without affecting other parts
of the program.

The central portion of our code is the ``main driver'' routine, which
is responsible for determining the flow of the program during
execution.  It determines which subroutines will be included in the
computation, what outputs will be required, when to write out data,
and at what point to stop the calculation.  The first task of the main
driver is to call a setup routine that reads in various input files,
sets up the coordinate grids, and either generates or reads in the
initial data, depending on the problem being run.  Once the initial
configuration has been determined control is passed back to the main
driver, which calls the physics driver to propagate the system forward
in time.  Whenever appropriate the main driver will call the routines
that output graphics, restart dumps, and other data.

The evolution of the physical system is carried out by the physics
driver, which coordinates the numerical integration of the
hydrodynamical equations with the calculation of the Newtonian
gravitational field and the gravitational radiation quantities.  The
hydrodynamics computation is performed by a large number of
subroutines, each devoted to a separate task such as the advection of
a particular quantity along a given coordinate direction.  The
gravitational potential calculation is separated into computing the
boundary conditions and solving Poisson's equation.

This modular structure permits the simple interchange of different
methods of implementing certain calculations, such as differencing the
advection terms, solving Poisson's equation, or computing the
gravitational radiation.  It further allows for changes in the
physical system such as specifying a different EOS or using a 1-D or
2-D grid to treat problems that do not require a full 3-D calculation.
One may choose to skip certain calculations, such as not solving
Poisson's equation for the case of a pure hydrodynamics problem
without gravity.  Finally, if it becomes desirable to include
additional physics, such as radiation transport,
magneto-hydrodynamics, or the effects of nuclear processes, this can
be done by adding new subroutines for computing these processes.

\subsection{Operator Splitting}

We use the technique of {\it operator splitting} (Wilson 1979; Bowers
\& Wilson 1991) to solve the hydrodynamical equations (\ref{mass-cyl})
- (\ref{euler-phi}).  In this method, terms are split off from the
original set of equations and are successively applied to update the
physical variables.  Thus, the full set of equations is not solved
simultaneously, and each variable is not updated due to the full set
of physical processes in a single operation.  Instead, each variable
is partially updated under the approximation that only a single
process is responsible, and each physical process is applied
sequentially to update that variable fully.  The order in which each
term is updated can be important for the accuracy and stability of the
code, and is generally determined by experience and experiment.  The
order used in our code is based on the accumulated experience of many
people with different numerical codes, and is thought to be
particularly good for treating shocks (Clancy 1989; Bowers \& Wilson
1991).

We first divide the solution of the hydrodynamical equations into a
``Lagrangian step'', including Lagrangian-like processes such as
acceleration and work done on a given fluid element, and a ``transport
step'', which includes both transport of material through the grid and
grid motion.  The Lagrangian step is performed first, with the
following sequence of operations:
\begin{tabbing}
{}~~~~~~\= (a)~\= update velocities due to pressure and gravitational
forces; \\
\> (b) update velocities and energy due to artificial viscosity, using the \\
\>\> updated velocities from step (a); \\
\> (c) update energy due to mechanical or compressional work, using the updated
quantities\\
\>\> from step (b).
\end{tabbing}
The transport step is then performed to update mass density, energy,
and velocities due to the transport of fluid across zone boundaries
and due to the changes in the zones arising from the grid motion.  The
transport step is broken into three parts, one for advection in each
of the three coordinate directions.  To minimize numerical artifacts,
the order in which the transport in each direction is performed is
varied with each cycle.  Thus, on the first cycle the radial advection
might be performed first, followed by the axial and angular
advections, while on the next cycle the order might be
axial-angular-radial, with the order changing on each successive cycle
until all six permutations have been exhausted.  At the end of the
transport calculation in each direction, a momentum conservation
condition is applied to update the velocities. For each successive
substep of the transport step, the input variables contain the updated
values from the previous substep.

\subsection{Lagrangian step}

The Lagrangian step begins by updating the terms containing the
accelerations due to gravity and pressure gradients.  For the $\varpi$
component of momentum, with the density held constant, the
acceleration part of (\ref{euler-r}) is simply
\begin{equation}
\rho\frac{\partial v}{\partial t} \; = \;
-\frac{\partial
P}{\partial\varpi}\:-\:\rho\frac{\partial\Phi}{\partial\varpi}.
\label{r-accel-d}
\end{equation}
Integrating this over the volume element $\Delta
V_j\equiv\varpi_j\Delta\varpi_j\Delta
z_{k+\frac{1}{2}}\Delta\varphi_{l+\frac{1}{2}}$ we obtain, after some
manipulation,\pagebreak[1] {\samepage\begin{eqnarray}
v^{p}_{j,k+\frac{1}{2},l+\frac{1}{2}} & = &
v^{n-\frac{1}{2}}_{j,k+\frac{1}{2},l+\frac{1}{2}}\:-
\frac{\Delta t^n}{\Delta\varpi_j} \;
\frac{p^{n}_{j+\frac{1}{2},k+\frac{1}{2},l+\frac{1}{2}}-p^{n}_{j-\frac{1}{2},k+\frac{1}{2},l+\frac{1}{2}}}
	{\rho^{n}_{j,k+\frac{1}{2},l+\frac{1}{2}}}
\nonumber\\
&&\;\;\;\;\;\;\;\;\;\;+\:\frac{\Delta t^n}{\Delta\varpi_j}
\left(\Phi^{n}_{j+\frac{1}{2},k+\frac{1}{2},l+\frac{1}{2}}-\Phi^{n}_{j-\frac{1}{2},k+\frac{1}{2},l+\frac{1}{2}}\right),
\label{r-accel}
\end{eqnarray}}\pagebreak[1]where $v^p$ is used to indicate that the velocity
has been {\it partially}
updated to the next time level $n+\frac{1}{2}$.  Here,
$\rho^{n}_{j,k+\frac{1}{2},l+\frac{1}{2}}$ is the radial edge centered
density defined by the volume averaging method given by
(\ref{rt-vol-avg}).  Similarly, (\ref{euler-z}) and (\ref{euler-phi})
become\pagebreak[1] {\samepage\begin{eqnarray}
u^{p}_{j+\frac{1}{2},k,l+\frac{1}{2}} & = &
u^{n-\frac{1}{2}}_{j+\frac{1}{2},k,l+\frac{1}{2}}\:-
\frac{\Delta t^n}{\Delta z_k} \;
\frac{p^{n}_{j+\frac{1}{2},k+\frac{1}{2},l+\frac{1}{2}}-p^{n}_{j+\frac{1}{2},k-\frac{1}{2},l+\frac{1}{2}}}
	{\rho^{n}_{j+\frac{1}{2},k,l+\frac{1}{2}}}
\nonumber\\
&&\;\;\;\;\;\;+\:\frac{\Delta t^n}{\Delta z_k}
\left(\Phi^{n}_{j+\frac{1}{2},k+\frac{1}{2},l+\frac{1}{2}}-\Phi^{n}_{j+\frac{1}{2},k-\frac{1}{2},l+\frac{1}{2}}\right)
\label{z-accel}
\end{eqnarray}}\pagebreak[1]
and\pagebreak[1] {\samepage\begin{eqnarray}
w^{p}_{j+\frac{1}{2},k+\frac{1}{2},l} & = &
w^{n-\frac{1}{2}}_{j+\frac{1}{2},k+\frac{1}{2},l}\:-
\frac{\Delta t^n}{\varpi_{j+\frac{1}{2}}\Delta\varphi_k} \;
\frac{p^{n}_{j+\frac{1}{2},k+\frac{1}{2},l+\frac{1}{2}}-p^{n}_{j+\frac{1}{2},k+\frac{1}{2},l-\frac{1}{2}}}
	{\rho^{n}_{j+\frac{1}{2},k+\frac{1}{2},l}}
\nonumber\\
&&\;\;\;\;\;\;+\:\frac{\Delta
t^n}{\varpi_{j+\frac{1}{2}}\Delta\varphi_k}
\left(\Phi^{n}_{j+\frac{1}{2},k+\frac{1}{2},l+\frac{1}{2}}-\Phi^{n}_{j+\frac{1}{2},k+\frac{1}{2},l-\frac{1}{2}}\right),
\label{phi-accel}
\end{eqnarray}}\pagebreak[1]where
the $z$ and $\varphi$ face centered densities are defined by
(\ref{z-vol-avg}) and (\ref{t-vol-avg}).

At this point in the code the effects of artificial viscosity are
included.  The discussion of these terms is deferred to the next
section, however, so that they may be treated in greater detail.

Next we consider the Lagrangian term for mechanical or compressional
work.  Separating this term out of (\ref{ener-cyl}) yields
\begin{equation}
\frac{\partial(\rho\varepsilon)}{\partial t} \; = \;
-\:P\left(\frac{1}{\varpi}\frac{\partial(\varpi v)}{\partial\varpi}
\,+\,\frac{\partial u}{\partial z}
\,+\,\frac{1}{\varpi}\frac{\partial w}{\partial\varphi}\right)
\;=\;-P\nabla\cdot\vec{v}.
\label{work}
\end{equation}
The finite difference version of this equation is
\begin{equation}
\frac{\Delta (\rho\varepsilon)}{\Delta t^{n+\frac{1}{2}}} \; = \;
\frac{(\rho\varepsilon)^C-(\rho\varepsilon)^n}{\Delta t^{n+\frac{1}{2}}}\; = \;
-\:P^{n+\frac{1}{2}}(\nabla\cdot\vec{v})^{Q},
\label{fdwork-1}
\end{equation}
where all quantities are evaluated at zone center, the common
subscripts $j\!+\!\frac{1}{2}$, $k\!+\!\frac{1}{2}$,
$l\!+\!\frac{1}{2}$ have been omitted for convenience, and we have
used the superscript $C$ and $Q$ to denote partial updates due to
compressional work and artificial viscosity, respectively.  The finite
difference velocity divergence that appears in~(\ref{fdwork-1}) is
given by\pagebreak[1] {\samepage\begin{eqnarray}\lefteqn{
(\nabla\cdot\vec{v})^{Q}_{j+\frac{1}{2},k+\frac{1}{2},l+\frac{1}{2}}
\; = \;
\frac
{\varpi_{j+1}v^{Q}_{j+1,k+\frac{1}{2},l+\frac{1}{2}}-\varpi_{j}v^{Q}_{j,k+\frac{1}{2},l+\frac{1}{2}}}
{\varpi_{j+\frac{1}{2}}\Delta\varpi_{j+\frac{1}{2}}}
\:+}\nonumber \\ &&\;\;\;\;\;\;\;\;
\frac{u^{Q}_{j+\frac{1}{2},k+1,l+\frac{1}{2}}-u^{Q}_{j+\frac{1}{2},k,l+\frac{1}{2}}}
{\Delta z_{k+\frac{1}{2}}}
+\frac{w^{Q}_{j+\frac{1}{2},k+\frac{1}{2},l+1}-w^{Q}_{j+\frac{1}{2},k+\frac{1}{2},l}}
{\varpi_{j+\frac{1}{2}}\Delta\varphi_{l+\frac{1}{2}}}.
\label{vdiv}
\end{eqnarray}}\pagebreak[1]The pressure that appears in (\ref{fdwork-1}) must
be evaluated at time step
$n+\frac{1}{2}$.  Since pressure is calculated from the EOS and the
density and energy are known only at time step $n$, we approximate
this by
\begin{displaymath}
P^{n+\frac{1}{2}}\;=\;P^n\:+\:\frac{1}{2}
\left(\frac{\partial P}{\partial\rho}\right)^n_{\varepsilon}\Delta\rho \:+\:
\frac{1}{2}\left(\frac{\partial
P}{\partial\varepsilon}\right)^n_{\rho}\Delta\varepsilon ,
\end{displaymath}
where the partial derivatives are evaluated using the known values of
$\rho$ and $\varepsilon$ at time step $n$.  Depending on the EOS being
used, an analytic form may be available for these partial derivatives,
or they may be interpolated from a table.  Using
$\Delta(\rho\varepsilon)=\rho^C\Delta\varepsilon +
\varepsilon^n\Delta\rho$, we convert (\ref{fdwork-1}) into an equation
for the change in internal energy,
\begin{equation}
\Delta\varepsilon\:=\:\varepsilon^C-\varepsilon^n\:=\:
- \; \frac{\varepsilon^n\Delta\rho + \left[P^n+\frac{1}{2}
\left(\frac{\partial P}{\partial\rho}\right)^n_{\varepsilon}\Delta\rho\right]
(\nabla\cdot\vec{v})^{Q}\Delta t^{n+\frac{1}{2}} }
{\rho^C+\frac{1}{2}\left(\frac{\partial
P}{\partial\varepsilon}\right)^n_{\rho} (\nabla\cdot\vec{v})^{Q}\Delta
t^{n+\frac{1}{2}} }.
\label{fdwork}
\end{equation}
We obtain $\Delta\rho$ and $\rho^C$ by using the fact that, since this
is a Lagrangian process, the density must obey the condition
\begin{equation}
\frac{\Delta\rho}{\rho^C}\;=\;\frac{\rho^C-\rho^n}{\rho^C}\;=\;
-(\nabla\cdot\vec{v})\Delta t^{n+\frac{1}{2}}.
\label{lag-rho}
\end{equation}
When these values are substituted into (\ref{fdwork}) along with
(\ref{vdiv}) and the partial derivatives of the EOS, we get an
explicit formula for updating the internal energy due to the effects
of mechanical work.  For the case of the ideal gas EOS
(\ref{ideal-gas}) we have
\begin{equation}
\varepsilon^C=\varepsilon^n\left\{1+(\nabla\cdot\vec{v})\Delta t
\frac{1-(\gamma -1)\left[1 + \frac{1}{2}(\nabla\cdot\vec{v})\Delta t\right]}
{1 + \frac{1}{2}(\gamma -1)\left[1 +
\frac{1}{2}(\nabla\cdot\vec{v})\Delta
t\right](\nabla\cdot\vec{v})\Delta t}\right\},
\label{int-mech}
\end{equation}
where the velocity divergence is evaluated using (\ref{vdiv}).

\subsection{Artificial Viscosity}\label{art-visc}

The effects of artificial viscosity are part of the Lagrangian step,
and these terms are actually updated in our code prior to the
treatment of mechanical work as noted above.  However, we have
separated the discussion of artificial viscosity from the rest of the
Lagrangian processes because it warrants a more thorough treatment.
Artificial viscosity is not a true physical process, but has been
developed to mimic the effects of physical processes in shock fronts
which occur on length scales much smaller than that of the numerical
grid.  The original concept was developed by Von Neumann \& Richtmeyer
(1949) and a modern discussion can be found in Bowers \& Wilson
(1991).

Artificial viscosity is designed to take the place of the microphysics
of true molecular viscosity in shock fronts.  By analogy to the
equations of molecular viscosity we write the artificial viscosity
terms in the momentum and energy equations in the form (Bowers \&
Wilson 1991):
\begin{equation}
\frac{\partial(\rho v^{a})}{\partial t}\;=\;-\nabla_{b}
\left(\rho Q^{ab}\right)
\label{artv1}
\end{equation}
and
\begin{equation}
\frac{\partial\varepsilon}{\partial t}\;=\;-Q^{ab}\frac{\partial v^a}{\partial
x^b}.
\label{artv2}
\end{equation}
The quantity $Q^{ab}$ is the tensor artificial viscosity defined by
\begin{eqnarray}
Q^{aa}&=&\left\{\begin{array}{ll} C_q\left(\Delta_a v^a\right)^2 &
\mbox{~~~~~~~~~if $\Delta_a v^a < 0$} \\ 0 & \mbox{~~~~~~~~~if
$\Delta_a v^a \geq 0$}
\end{array}\right. \mbox{(no summation)} \label{avtensor}\\
Q^{ab}&=&C_{shear}\left(\Delta_b v^a\right)\left|\Delta_b v^a\right|
\mbox{~~~if $a \ne b$},
\nonumber
\end{eqnarray}
where $C_q$ and $C_{shear}$ are constants that specify the amount of
compressional and shear viscosity, respectively, and are given as
input to the program. We use the notation
\begin{displaymath}
\Delta_a \equiv \Delta x^a \frac{\partial}{\partial x^a}
\mbox{~~~~~(no summation)}.
\end{displaymath}
The definition of the diagonal terms $Q^{aa}$ in (\ref{avtensor})
ensures that artificial viscosity is present under circumstances of
compression but not expansion.  Currently we are not including shear
artificial viscosity in our code, so we take $C_{shear} = 0$ and limit
our attention to the compressional terms.

When the artificial viscosity is placed on the numerical grid the
compressional terms are zone centered.  Thus the finite difference
version of the radial component of (\ref{avtensor}) is
\begin{equation}
Q^{\varpi\varpi}_{j+\frac{1}{2},k+\frac{1}{2},l+\frac{1}{2}}\;=\;C_q
\left(v^p_{j+1,k+\frac{1}{2},l+\frac{1}{2}}-v^p_{j,k+\frac{1}{2},l+\frac{1}{2}}\right)^2,
\label{Qrr}
\end{equation}
assuming that $\Delta v < 0$.  Note that the artificial viscosity
tensor is calculated using the velocity that has been partially
updated by the pressure and potential gradients.  The other diagonal
components, again assuming that $\Delta u < 0$ and $\Delta w < 0$, are
given by
\begin{equation}
Q^{zz}_{j+\frac{1}{2},k+\frac{1}{2},l+\frac{1}{2}}\;=\;C_q
\left(v^p_{j+\frac{1}{2},k+1,l+\frac{1}{2}}-v^p_{j+\frac{1}{2},k,l+\frac{1}{2}}\right)^2
\end{equation}
and
\begin{equation}
Q^{\varphi\varphi}_{j+\frac{1}{2},k+\frac{1}{2},l+\frac{1}{2}}\;=\;C_q
\left(w^p_{j+\frac{1}{2},k+\frac{1}{2},l+1}-w^p_{j+\frac{1}{2},k+\frac{1}{2},l}\right)^2.
\label{Qtt}
\end{equation}

With these expressions the acceleration terms (\ref{artv1}), including
only the compressional viscosity, can easily be differenced to obtain
\begin{equation}
v^Q_j\;=\;v^p_j-\frac{\Delta t^n}{\rho^n_j} \;
\frac{\left(\varpi\rho Q^{\varpi\varpi}\right)_{j+\frac{1}{2}}-
\left(\varpi\rho
Q^{\varpi\varpi}\right)_{j-\frac{1}{2}}}{\varpi_j\Delta\varpi_j},
\label{artvv}
\end{equation}
\begin{equation}
u^Q_k\;=\;u^p_k-\frac{\Delta t^n}{\rho^n_k} \;
\frac{\left(\rho Q^{zz}\right)_{k+\frac{1}{2}}-
\left(\rho Q^{zz}\right)_{k-\frac{1}{2}}}{\Delta z_k},
\label{artvu}
\end{equation}
and
\begin{equation}
w^Q_l\;=\;w^p_l-\frac{\Delta t^n}{\rho^n_l} \;
\frac{\left(\rho Q^{\varphi\varphi}\right)_{l+\frac{1}{2}}-
\left(\rho Q^{\varphi\varphi}\right)_{l-\frac{1}{2}}}{\varpi\Delta\varphi_l}.
\label{artvw}
\end{equation}
Here we have suppressed the common subscripts
$j+\frac{1}{2},k+\frac{1}{2},l+\frac{1}{2}$ wherever they appear and
have used the superscript $Q$ to indicate the partial update due to
artificial viscosity.

Finally, the shock heating from (\ref{artv2}) is
\begin{equation}
\varepsilon^Q=\varepsilon^n-\Delta t^{n+\frac{1}{2}}\left[
Q^{\varpi\varpi}\frac{v^Q_{j+1}-v^Q_j}{\Delta\varpi_{j+\frac{1}{2}}} +
Q^{\varphi\varphi}\frac{w^Q_{l+1}-w^Q_l}{\varpi\Delta\varphi_{l+\frac{1}{2}}}
+ Q^{zz}\frac{u^Q_{k+1}-u^Q_k}{\Delta z_{k+\frac{1}{2}}} \right].
\end{equation}
Again, the common subscripts
$j+\frac{1}{2},k+\frac{1}{2},l+\frac{1}{2}$ have been suppressed in
each term.  Note that although the velocities used in the shock
heating have already been updated by (\ref{artvv})--(\ref{artvw}), the
artificial viscosity tensor is the same one calculated previously and
used to update the velocities.

\subsection{Transport Step}\label{sec-trans}

The transport step includes terms involving advection and the change
in zone size.  It is divided into three parts, for transport along
each of the three coordinate directions.  To minimize numerical
effects, the order in which these substeps are performed is varied on
each cycle through all possible permutations.  We use a second order
monotonic advection scheme developed by LeBlanc (Clancy 1989; Bowers
\& Wilson 1991), and the consistent advection scheme of Norman and
Winkler (1986) for angular momentum transport.

First, consider the transport of mass density in the $z$-direction.
The relevant transport terms from (\ref{mass-cyl}) are
\begin{equation}
\frac{\partial\rho}{\partial t}\;=\;
-\:\frac{\partial(\rho u_r)}{\partial z}
\:-\:\rho \frac{\partial u_g}{\partial z}.
\label{ztrans-d}
\end{equation}
Integrating this equation over a zone volume and rearranging terms
gives
\begin{equation}
\rho^{Az}_{k+\frac{1}{2}}\;=\;\rho^{n}_{k+\frac{1}{2}}\left\{1-
\frac{\Delta t^{n+\frac{1}{2}}}{\Delta z_{k+\frac{1}{2}}}
\left[{u_g}_{k+1}-{u_g}_{k}\right]\right\}
-\frac{\Delta t^{n+\frac{1}{2}}}{\Delta z_{k+\frac{1}{2}}}
\left[\left(\rho^*u_r\right)_{k+1}-\left(\rho^*u_r\right)_{k}\right],
\label{zadv-1}
\end{equation}
where we have suppressed the indices $j+\frac{1}{2},l+\frac{1}{2}$,
and the superscript $Az$ is used to indicate that the density has been
partially updated by advection in the $z$-direction.  The quantity
$\rho^*_k$ represents the density interpolated to the zone face; this
interpolation is performed using a monotonic scheme as described in
\S~\ref{sec-mono} below.  The first term on the right hand side of
(\ref{zadv-1}) can be shown to be a first order approximation to the
change in density due to the change in the cell volume caused by the
motion of the $z$-grid.  Since the second term, which specifies the
change in density due to the transport of matter across the zone
boundaries, will be given to second order by the monotonic
interpolation scheme specified below, the first order term limits the
accuracy of the calculation when the grid velocity is non-zero.
However, if the grid position at the updated time step, $n\!+\!1$, is
known, we can replace this approximation by the exact expression to
give
\begin{equation}
\rho^{Az}_{k+\frac{1}{2}}\;=\;\rho^{n}_{k+\frac{1}{2}}
\frac{\Delta z^{n}_{k+\frac{1}{2}}}{\Delta z^{n+1}_{k+\frac{1}{2}}}
-\frac{\Delta t^{n+\frac{1}{2}}}{\Delta z^{n+1}_{k+\frac{1}{2}}}
\left[\left(\rho^*u_r\right)_{k+1}-\left(\rho^*u_r\right)_{k}\right],
\label{zadv-d}
\end{equation}
where we have also used the updated zone spacing in the second term.
Since the effects of the change in grid spacing are treated exactly in
(\ref{zadv-d}), the accuracy is determined by the transport term,
which is second order.  This gives a higher degree of accuracy than
(\ref{zadv-1}) when the moving grid is used (and the same accuracy
when the grid is stationary), and therefore we use (\ref{zadv-d}) in
our code.

The energy transport in the $z$-directon can be treated in similar
fashion, yielding
\begin{equation}
(\rho\varepsilon)^{Az}_{k+\frac{1}{2}}\;=\;(\rho\varepsilon)^{C}_{k+\frac{1}{2}}
\; \frac{\Delta z^{n}_{k+\frac{1}{2}}}{\Delta z^{n+1}_{k+\frac{1}{2}}}
-\frac{\Delta t^{n+\frac{1}{2}}}{\Delta z^{n+1}_{k+\frac{1}{2}}}
\left[\left(\rho^*\varepsilon^*u_r\right)_{k+1}-
\left(\rho^*\varepsilon^*u_r\right)_{k}\right],
\label{zadv-e}
\end{equation}
where $\varepsilon^*$ is calculated using a monotonic interpolation
scheme, and we have again suppressed the common subscripts
$j+\frac{1}{2},l+\frac{1}{2}$.  The specific internal energy at the
updated time level is then
\begin{equation}
\varepsilon^{Az}\;=\;\frac{(\rho\varepsilon)^{Az}}{\rho^{Az}},
\label{int-Az}\end{equation}
where $\rho^{Az}$ is defined in (\ref{zadv-d}).  The term
$(\rho\epsilon)^C$ in~(\ref{zadv-e}) involves the internal energy
$\epsilon^C$, which was updated for mechanical work in
equation~(\ref{int-mech}) in the Lagrangian step.  This is given by
\begin{equation}
(\rho\epsilon)^C = \rho^C \epsilon^C,
\label{rho-mech} \end{equation}
where $\rho^C$ is the density updated under Lagrangian conditions for
mechanical work similar to those used to get $\epsilon^C$.  To obtain
$\rho^C$ we apply the Lagrangian density condition (\ref{lag-rho}),
but include only the effects of axial velocity; this gives
\begin{equation}
\rho^C_{k+\frac{1}{2}}\;=\;\frac{\rho^n_{k+\frac{1}{2}}}{1+\left(\nabla\cdot\vec{v}_z
\right)^Q_{k+\frac{1}{2}}\Delta t^{n+\frac{1}{2}}},
\label{rhoc}
\end{equation}
where
\begin{displaymath}
\left(\nabla\cdot\vec{v}_z\right)^Q_{k+\frac{1}{2}}\;=\;
\frac{u^Q_{k+1}-u^Q_k}{\Delta z_{k+\frac{1}{2}}},
\end{displaymath}
and the superscript $Q$ indicates that the velocity has been updated
by artificial viscosity.

A similar procedure is used to transport the radial and axial momentum
in the $z$-direction in (\ref{euler-r}) and (\ref{euler-z}), yielding
\begin{equation}
v^{Az}_{k+\frac{1}{2}}=v^{Q}_{k+\frac{1}{2}}
\frac{\Delta z^{n}_{k+\frac{1}{2}}}{\Delta z^{n+1}_{k+\frac{1}{2}}}
-\frac{\Delta t^{n+\frac{1}{2}}}{\Delta z^{n+1}_{k+\frac{1}{2}}}
\left[\frac{\left(\rho^*v^*u_r\right)_{k+1}-
\left(\rho^*v^*u_r\right)_{k}}{\rho^n_{k+\frac{1}{2}}}\right],
\label{zadv-v}
\end{equation}
where the common subscripts $j,l+\frac{1}{2}$ have been suppressed,
and
\begin{equation}
u^{Az}_{k}\;=\;u^{Q}_{k}
\frac{\Delta z^{n}_{k}}{\Delta z^{n+1}_{k}}
-\frac{\Delta t^{n+\frac{1}{2}}}{\Delta z^{n+1}_{k}}
\left[\frac{\left(\rho^*u^*u_r\right)_{k+\frac{1}{2}}-
\left(\rho^*u^*u_r\right)_{k-\frac{1}{2}}}{\rho^n_{k}}\right],
\label{zadv-u}
\end{equation}
where the common subscripts $j+\frac{1}{2},l+\frac{1}{2}$ have been
suppressed.  Several of the quantities in these equations are not
evaluated at their usual positions on the grid, and these values are
obtained by volume averaging as described in \S~\ref{sec-grid-cen}
above.  In particular, the edge centered velocity
$\left.u_r\right._{j,k,l+\frac{1}{2}}$ in (\ref{zadv-v}) and the zone
centered velocity
$\left.u_r\right._{j+\frac{1}{2},k+\frac{1}{2},l+\frac{1}{2}}$ in
(\ref{zadv-u}) must be obtained from the velocity field
$\left.u_r\right._{j+\frac{1}{2},k,l+\frac{1}{2}}$ before the
monotonic scheme is applied to obtain $v^*$ and $u^*$ in each
equation.  The densities $\rho^*_{j,k,l+\frac{1}{2}}$ and
$\rho^*_{j+\frac{1}{2},k+\frac{1}{2},l+\frac{1}{2}}$ are obtained by a
volume average of the monotonic density
$\rho^*_{j+\frac{1}{2},k,l+\frac{1}{2}}$ that was calculated in
(\ref{zadv-d}).

The advection of angular momentum presents special problems in finite
difference codes, and will be discussed more fully in
\S~\ref{sec-mono} below.  At this point we simply state that it is
advantageous to perform the advection in terms of the specific angular
momentum, $J\equiv\varpi w$.  When placed on the finite difference
grid, this becomes
\begin{equation}
J_{j+\frac{1}{2},k+\frac{1}{2},l} = w_{j+\frac{1}{2},k+\frac{1}{2},l}
\frac{\varpi^2_{j+1}+\varpi_{j+1}\varpi_j+\varpi^2_j}{3\varpi_{j+\frac{1}{2}}}.
\label{spec-ang-mom}
\end{equation}
The $z$-transport terms from (\ref{euler-phi}) can be rewritten in
terms of the specific angular momentum as
\begin{equation}
\frac{\partial(\rho J)}{\partial t}\;  = \;
-\frac{\partial(\rho J u_r)}{\partial z} -\:\rho J\frac{\partial
u_g}{\partial z} .
\end{equation}
In finite difference form this becomes
\begin{equation}
J^{Az}_{k+\frac{1}{2}}=J^{Q}_{k+\frac{1}{2}} \;
\frac{\Delta z^{n}_{k+\frac{1}{2}}}{\Delta z^{n+1}_{k+\frac{1}{2}}}
-\frac{\Delta t^{n+\frac{1}{2}}}{\Delta z^{n+1}_{k+\frac{1}{2}}}
\left[\frac{\left(\rho^*\hat{J}^*u_r\right)_{k+1}-
\left(\rho^*\hat{J}^*u_r\right)_{k}}{\rho^n_{k+\frac{1}{2}}}\right],
\label{zadv-w}
\end{equation}
where we have again suppressed the indices $j+\frac{1}{2},l$.  The
symbol $\hat{J}^{*}$ is used to indicate that we have used a
``smooth'' value of $J$ in the monotonic interpolation scheme for the
transport terms; see equation~(\ref{smooth}) below.  As before, we
obtain the edge centered relative velocity
$\left.u_r\right._{j+\frac{1}{2},k,l}$ by volume averaging before
applying the monotonic interpolation scheme to the angular quantities,
and get the density $\rho^*_{j+\frac{1}{2},k,l}$ from a volume average
of the interpolated density in (\ref{zadv-d}).

During the momentum advection the density has been held constant at
time level $n$.  In order to ensure momentum conservation as the
density is updated to time step $n+1$ we must update the velocity by
\begin{equation}
\vec{v}\,^{n+1}\;=\;\vec{v}\,^n\frac{\rho^n}{\rho^{n+1}},
\label{mocon}
\end{equation}
where $\rho$ is evaluated at the appropriate grid location for each of
the velocity components.  In component form, this gives
\begin{displaymath}
v^{n+1}_{j,k+\frac{1}{2},l+\frac{1}{2}}\;=\;\left[v^n
\frac{\rho^n}{\rho^{n+1}}\right]_{j,k+\frac{1}{2},l+\frac{1}{2}},
\end{displaymath}
\begin{equation}
u^{n+1}_{j+\frac{1}{2},k,l+\frac{1}{2}}\;=\;\left[u^n
\frac{\rho^n}{\rho^{n+1}}\right]_{j+\frac{1}{2},k,l+\frac{1}{2}},
\label{mocon-cpts}
\end{equation}
and
\begin{displaymath}
w^{n+1}_{j+\frac{1}{2},k+\frac{1}{2},l}\;=\;\left[w^n
\frac{\rho^n}{\rho^{n+1}}\right]_{j+\frac{1}{2},k+\frac{1}{2},l},
\end{displaymath}
where the face centered densities are again calculated using volume
averaging.  This update completes the transport in the axial
direction.

The advection in the radial and angular directions is accomplished in
a similar fashion, with results as follows.

For radial advection:
\begin{equation}
\rho^{A\varpi}_{j+\frac{1}{2}}=\rho^{n}_{j+\frac{1}{2}}
\frac{(\varpi\Delta\varpi)^{n}_{j+\frac{1}{2}}}{(\varpi\Delta\varpi)^{n+1}_{j+\frac{1}{2}}}
-\frac{\Delta
t^{n+\frac{1}{2}}}{(\varpi\Delta\varpi)^{n+1}_{j+\frac{1}{2}}}
\left[\left(\varpi^*\rho^*v_r\right)_{j+1}-\left(\varpi^*\rho^*v_r\right)_{j}\right],
\label{radv-d}
\end{equation}
\begin{equation}
(\rho\varepsilon)^{A\varpi}_{j+\frac{1}{2}}=(\rho\varepsilon)^{C}_{j+\frac{1}{2}}
\frac{(\varpi\Delta\varpi)^{n}_{j+\frac{1}{2}}}{(\varpi\Delta\varpi)^{n+1}_{j+\frac{1}{2}}}
-\frac{\Delta
t^{n+\frac{1}{2}}}{(\varpi\Delta\varpi)^{n+1}_{j+\frac{1}{2}}}
\left[\left(\varpi^*\rho^*\varepsilon^*v_r\right)_{j+1}-
\left(\varpi^*\rho^*\varepsilon^*v_r\right)_{j}\right],
\label{radv-e}
\end{equation}
\begin{equation}
v^{A\varpi}_{j}=v^{Q}_{j}
\frac{(\varpi\Delta\varpi)^{n}_{j}}{(\varpi\Delta\varpi)^{n+1}_{j}}
-\frac{\Delta
t^{n+\frac{1}{2}}}{(\varpi\Delta\varpi)^{n+1}_{j+\frac{1}{2}}}
\left[\frac{\left(\varpi^*\rho^*v^*v_r\right)_{j+\frac{1}{2}}-
\left(\varpi^*\rho^*v^*v_r\right)_{j-\frac{1}{2}}}{\rho^n_{j}}\right],
\label{radv-v}
\end{equation}
\begin{equation}
u^{A\varpi}_{j+\frac{1}{2},k}=u^{Q}_{j+\frac{1}{2},k}
\frac{(\varpi\Delta\varpi)^{n}_{j+\frac{1}{2}}}{(\varpi\Delta\varpi)^{n+1}_{j+\frac{1}{2}}}
-\frac{\Delta
t^{n+\frac{1}{2}}}{(\varpi\Delta\varpi)^{n+1}_{j+\frac{1}{2}}}
\left[\frac{\left(\varpi^*\rho^*u^*v_r\right)_{j+1}-
\left(\varpi^*\rho^*u^*v_r\right)_{j}}{\rho^n_{j+\frac{1}{2}}}\right]_k,
\label{radv-u}
\end{equation}
and
\begin{equation}
J^{A\varpi}_{j+\frac{1}{2},l}=J^{Q}_{j+\frac{1}{2},l}
\frac{(\varpi\Delta\varpi)^{n}_{j+\frac{1}{2}}}{(\varpi\Delta\varpi)^{n+1}_{j+\frac{1}{2}}}
-\frac{\Delta
t^{n+\frac{1}{2}}}{(\varpi\Delta\varpi)^{n+1}_{j+\frac{1}{2}}}
\left[\frac{\left(\varpi^*\rho^*\hat{J}^*v_r\right)_{j+1}-
\left(\varpi^*\rho^*\hat{J}^*v_r\right)_{j}}{\rho^n_{j+\frac{1}{2}}}\right]_l.
\label{radv-w}
\end{equation}
Here we have suppressed the indices $k+\frac{1}{2},l+\frac{1}{2}$
throughout.  The coordinate
\begin{equation}
\varpi^*_j=\varpi_j - \textstyle{\frac{1}{2}} {v_r}_j\Delta t
\label{varpi-star}
\end{equation}
 is the radial position at which the monotonically interpolated
quantities are evaluated, as discussed in the next section.  The
velocities ${v_r}_{j+\frac{1}{2}}$, ${v_r}_{j,k}$ and ${v_r}_{j,l}$ in
(\ref{radv-v}), (\ref{radv-u}) and (\ref{radv-w}), respectively, are
obtained by volume averaging before the monotonic interpolation scheme
is applied, and the densities $\rho^*_{j+\frac{1}{2}}$, $\rho^*_{j,k}$
and $\rho^*_{j,l}$ are obtained by volume averaging the density
$\rho^*_j$ that results from monotonic interpolation in the radial
direction.  The specific internal energy $\epsilon^{A\varpi}$ is
obtained from (\ref{radv-e}) by using (\ref{rhoc}), with the velocity
divergence term replaced with the radial velocity divergence
\begin{displaymath}
\left(\nabla\cdot\vec{v}_\varpi\right)^Q_{k+\frac{1}{2}}\;=\;
\frac{\varpi_{j+1}v^Q_{j+1}-\varpi_{j}v^Q_j}{\varpi_{j+\frac{1}{2}}\Delta\varpi_{j+\frac{1}{2}}}.
\end{displaymath}

For angular advection, with the indices $j+\frac{1}{2},k+\frac{1}{2}$
suppressed throughout:
\begin{equation}
\rho^{A\varphi}_{l+\frac{1}{2}}=\rho^{n}_{l+\frac{1}{2}}
\frac{\Delta\varphi^{n}_{l+\frac{1}{2}}}{\Delta\varphi^{n+1}_{l+\frac{1}{2}}}
-\frac{\Delta
t^{n+\frac{1}{2}}}{\varpi\Delta\varphi^{n+1}_{l+\frac{1}{2}}}
\left[\left(\rho^*w_r\right)_{l+1}-\left(\rho^*w_r\right)_{l}\right],
\label{wadv-d}
\end{equation}
\begin{equation}
(\rho\varepsilon)^{A\varphi}_{l+\frac{1}{2}}=(\rho\varepsilon)^{C}_{l+\frac{1}{2}}
\frac{\Delta\varphi^{n}_{l+\frac{1}{2}}}{\Delta\varphi^{n+1}_{l+\frac{1}{2}}}
-\frac{\Delta
t^{n+\frac{1}{2}}}{\varpi\Delta\varphi^{n+1}_{l+\frac{1}{2}}}
\left[\left(\rho^*\varepsilon^*w_r\right)_{l+1}-
\left(\rho^*\varepsilon^*w_r\right)_{l}\right],
\label{wadv-e}
\end{equation}
\begin{equation}
v^{A\varphi}_{j,l+\frac{1}{2}}=v^{Q}_{j}
\frac{\Delta\varphi^{n}_{l+\frac{1}{2}}}{\Delta\varphi^{n+1}_{l+\frac{1}{2}}}
-\frac{\Delta
t^{n+\frac{1}{2}}}{\varpi_{j}\Delta\varphi^{n+1}_{l+\frac{1}{2}}}
\left[\frac{\left(\rho^*v^*w_r\right)_{l+1}-
\left(\rho^*v^*w_r\right)_{l}}{\rho^n_{l+\frac{1}{2}}}\right]_j,
\label{wadv-v}
\end{equation}
\begin{equation}
u^{A\varphi}_{l+\frac{1}{2},k}=u^{Q}_{l+\frac{1}{2},k}
\frac{\Delta\varphi^{n}_{l+\frac{1}{2}}}{\Delta\varphi^{n+1}_{l+\frac{1}{2}}}
-\frac{\Delta
t^{n+\frac{1}{2}}}{\varpi\Delta\varphi^{n+1}_{l+\frac{1}{2}}}
\left[\frac{\left(\rho^*u^*w_r\right)_{l+1}-
\left(\rho^*u^*w_r\right)_{l}}{\rho^n_{l+\frac{1}{2}}}\right]_k,
\label{wadv-u}
\end{equation}
and
\begin{equation}
J^{A\varphi}_{l}=J^{Q}_{l}
\frac{\Delta\varphi^{n}_{l}}{\Delta\varphi^{n+1}_{l}}
-\frac{\Delta t^{n+\frac{1}{2}}}{\varpi\Delta\varphi^{n+1}_{l}}
\left[\frac{\left(\rho^*\hat{J}^*w_r\right)_{l+\frac{1}{2}}-
\left(\rho^*\hat{J}^*w_r\right)_{l-\frac{1}{2}}}{\rho^n_{l}}\right].
\label{wadv-w}
\end{equation}
We again evaluate ${w_r}_{j,l}$, ${w_r}_{l,k}$ and
${w_r}_{l+\frac{1}{2}}$ by volume averaging before applying the
monotonic interpolation and obtain $\rho^*_{j,l}$, $\rho^*_{l,k}$ and
$\rho^*_{l+\frac{1}{2}}$ by volume averaging the density $\rho^*_l$,
which is monotonically interpolated in the angular direction.  As
before, we get $\varepsilon^{A\varphi}$ from (\ref{wadv-e}) and
(\ref{rhoc}) with the velocity divergence term now replaced with its
angular analog,
\begin{displaymath}
\left(\nabla\cdot\vec{v}_\varphi\right)^Q_{l+\frac{1}{2}}\;=\;
\frac{w^Q_{l+1}-w^Q_l}{\Delta \varphi_{l+\frac{1}{2}}}.
\end{displaymath}

As noted above, the order in which the advection is carried out along
each of the coordinate directions is varied with each time step
through all six permutations.  This is done to minimize any
non-physical effects that might arise from carrying out the advection
in the same order on each cycle.  The momentum conservation
calculation (\ref{mocon})--(\ref{mocon-cpts}) is performed at the end
of the transport in each direction.  Caution should thus be exercised
in interpreting each of the partially updated time levels occurring in
(\ref{zadv-1})--(\ref{wadv-w}).  For example, the values of $\rho^n$,
$\varepsilon^C$, and $v^Q$ may refer to the values of those variables
after advection along one or more directions rather than at the end of
the Lagrangian step. The full time level update $n+1$ indicated in
(\ref{mocon})--(\ref{mocon-cpts}) is only achieved after advection has
been completed in all three directions.

\subsection{Treatment of Advection}\label{sec-mono}

In transporting physical quantities from one zone to another it is
necessary to obtain an accurate value for the quantity crossing the
zone face.  This is accomplished by using a monotonic interpolation
scheme from the zone center to the zone face.  The method that we are
currently using is a second order accurate scheme developed by LeBlanc
(Clancy 1989; Bowers \& Wilson 1991), although the option currently
exists in the code to use either donor cell advection or a monotonic
scheme developed by Barton (see Centrella \& Wilson 1984).

The simplest advection scheme is the donor cell method, in which the
density of the material that flows from one cell to another is simply
taken to be the density in the cell that it is flowing out of.  For
example, if we are considering mass advection in the radial direction
between cell $j-\frac{1}{2}$ and cell $j+\frac{1}{2}$, then the value
of the density would be taken as $\rho^*_j=\rho_{j-\frac{1}{2}}$ if
${v_r}_{j}>0$ and $\rho^*_j=\rho_{j+\frac{1}{2}}$ if ${v_r}_{j}<0$.
The total mass transferred from cell $j-\frac{1}{2}$ to cell
$j+\frac{1}{2}$ would be $\Delta m = \rho^*_j\varpi_j\Delta
z\Delta\varphi {v_r}_{j}\Delta t$.  This method is only accurate to
first order and produces large numerical diffusion (Bowers \& Wilson
1991).

In order to achieve higher accuracy, it is preferable to interpolate
zone centered quantities to face centers using a method that preserves
monotonicity in the quantity being advected (van Leer 1974; Bowers \&
Wilson 1991).  Monotonic advection prevents zones from being evacuated
and eliminates unstable zone to zone oscillations; see Hawley, Smarr
and Wilson (1984b) for a study of various second order monotonic
schemes.  We have found the LeBlanc method to produce the best results
in test problems.

The LeBlanc interpolation scheme is illustrated in
Figure~\ref{monofig} for the advection of density in the
$z$-direction; a similar treatment applies to the advection of all
quantities along any direction.  The purpose of the interpolation is
to determine the value of the density at the zone face $z_k$
(actually, near the zone face, as described below).  We use the
density in zone $z_{k+\frac{1}{2}}$ and the two neighboring zones to
get the slope of the density function, $S_{k+\frac{1}{2}}$, in that
zone, and then use that slope to interpolate the density to the zone
face.  To obtain $S_{k+\frac{1}{2}}$ we first calculate the three
slopes $S_L,S_R$, and $S_2$ as shown in Figure~\ref{monofig}.  We
assume that the densities $\rho_{k-\frac{1}{2}}$ and
$\rho_{k+\frac{3}{2}}$ are constant in the cells $k-\frac{1}{2}$ and
$k+\frac{3}{2}$, respectively.  The ``left'' slope $S_L$ is then
calculated by using $\rho_{k-\frac{1}{2}}$ at the zone face $z_k$ and
assuming that the density varies linearly from that value to the value
$\rho_{k+\frac{1}{2}}$, as shown in Figure~\ref{monofig}.  This gives
\begin{displaymath}
S_L=\frac{\rho_{k+\frac{1}{2}}-\rho_{k-\frac{1}{2}}}{\Delta
z_{k+\frac{1}{2}}/2}.
\end{displaymath}
  A similar calculation using $\rho_{k+\frac{3}{2}}$ at the zone face
$z_{k+1}$ gives the ``right'' slope $S_R$, where
\begin{displaymath}
S_R=\frac{\rho_{k+\frac{3}{2}}-\rho_{k+\frac{1}{2}}}{\Delta
z_{k+\frac{1}{2}}/2}.
\end{displaymath}
The third slope, $S_2$, is calculated by first linearly interpolating
the density $\rho_{k-\frac{1}{2}}$ from the zone center to the zone
face $z_k$, and the density $\rho_{k+\frac{3}{2}}$ to the zone face
$z_{k+1}$. The slope between those values gives $S_2$, which is
\begin{displaymath}
S_2 = \frac{1}{\Delta z_{k+\frac{1}{2}}}\left(
\frac{\rho_{k+\frac{1}{2}}\Delta z_{k+\frac{3}{2}}+\rho_{k+\frac{3}{2}}\Delta
z_{k+\frac{1}{2}}}
	{\Delta z_{k+\frac{3}{2}}+\Delta z_{k+\frac{1}{2}}}
\;-\; \frac{\rho_{k-\frac{1}{2}}\Delta
z_{k+\frac{1}{2}}+\rho_{k+\frac{1}{2}}\Delta z_{k-\frac{1}{2}}}
	{\Delta z_{k+\frac{1}{2}}+\Delta z_{k-\frac{1}{2}}} \right).
\end{displaymath}
The slope $S_{k+\frac{1}{2}}$ is then chosen to be the one from the
set $\{S_L,S_R,S_2\}$ with the minimum absolute value, or zero if any
two of the slopes have opposite sign.  The interpolated density is
calculated using this slope and is given by
\begin{equation}
\rho^*_k \;=\; \left\{\begin{array}{ll}
\rho_{k-\frac{1}{2}}+\frac{1}{2} S_{k-\frac{1}{2}}\left(\Delta
z_{k-\frac{1}{2}}-u_k\Delta t\right) & u_k>0 \\
\rho_{k+\frac{1}{2}}-\frac{1}{2} S_{k+\frac{1}{2}}\left(\Delta
z_{k+\frac{1}{2}}+u_k\Delta t\right) & u_k\leq 0
\end{array}\right.   ;
\label{eqmono}
\end{equation}
this is the density $\rho^*_k$ that appears in the axial mass
advection equation (\ref{zadv-d}) and the other axial advection
equations.  Note that the position at which the density is evaluated
is not the zone edge itself, but rather a distance $\frac{1}{2}
u_k\Delta t$ ``upwind'' from the zone edge $z_k$.  This is the
position of the center of the volume that will be transported across
that zone boundary in the time step $\Delta t$.  To simplify this
illustration, we have ignored grid motion; this may be included by
replacing the velocity $u$ with the relative velocity $u_r=u-u_g$ in
(\ref{eqmono}) and in Figure~\ref{monofig}.

A similar procedure is applied to interpolate each quantity to be
advected, and the use of a superscript $*$ in the transport equations
of \S~\ref{sec-trans} indicates quantities that must be evaluated
using monotonic interpolation.  Note that the coordinate $\varpi^*_j$,
defined in equation~(\ref{varpi-star}), appears in the radial
transport equations.  This is the position at which the variable is
evaluated in the monotonic interpolation scheme for transport in the
$\varpi$-direction.

In developing the advection equations above, we stated that it was
advantageous to recast Euler's equations in terms of specific angular
momentum, $J=\varpi w$, and that the angular velocity components are
updated using this formulation.  The reason for this will now be
explained.

The advection of angular momentum presents special problems in finite
difference codes.  Numerical effects can cause a gradual loss of total
angular momentum or a non-physical diffusion of angular momentum
outwards from the rotation axis.  These effects can be minimized by
employing a consistent advection scheme for angular momentum advection
(Norman \& Winkler 1986; Norman, Wilson, \& Barton 1980). In this
method, we must calculate the specific angular momentum $J$ and the
angular velocity $\Omega=w/\varpi$, in addition to the linear velocity
in the angular direction $w$, which is the variable used in the code.
On the numerical grid $J$ is given by (\ref{spec-ang-mom}) and the
angular velocity is obtained from
\begin{equation}
\Omega_{j+\frac{1}{2},k+\frac{1}{2},l} = J_{j+\frac{1}{2},k+\frac{1}{2},l} \;
\frac{2}{\varpi^2_{j+1}+\varpi^2_{j}} .
\end{equation}
The key to the consistent advection method is to perform advection
based on the smoothest of the three angular quantities, $w$, $J$ and
$\Omega$.  Each of these quantities is interpolated to the zone
boundary using the LeBlanc monotonic scheme described above.  For the
case of advection in the axial direction the ``smooth'' specific
angular momentum is defined to be
\begin{equation}
\hat{J}^*_{k}=\left\{\begin{array}{ll}
\Omega_{k}^*\varpi^2, & \mbox{\rule[-5pt]{0pt}{25pt}if
${\displaystyle\left|\frac{\Omega_{k+\frac{1}{2}}-\Omega_{k-\frac{1}{2}}}{\Omega^*_k}
\right|}$ is smallest} \\
w_{k}^*\varpi, & \mbox{\rule[-5pt]{0pt}{25pt}if
${\displaystyle\left|\frac{w_{k+\frac{1}{2}}-w_{k-\frac{1}{2}}}{w^*_k}
\right|}$ is smallest} \\
J_{k}^*, & \mbox{\rule[-5pt]{0pt}{25pt}if
${\displaystyle\left|\frac{J_{k+\frac{1}{2}}-J_{k-\frac{1}{2}}}{J^*_k}
\right|}$ is smallest}
\end{array}\right. ,
\label{smooth}
\end{equation}
where we have suppressed the indices $j+\frac{1}{2},l$.  This smooth
quantity $\hat{J}^*_k$ is the one that appears in the angular momentum
transport equations (\ref{zadv-w}), (\ref{radv-w}) and (\ref{wadv-w}).
The new linear velocity in the angular direction is obtained by
inverting (\ref{spec-ang-mom}) at the end of the specific angular
momentum transport.

The equation for obtaining the smooth specific angular momentum in the
angular direction is completely analogous to (\ref{smooth}), with an
appropriate change of indices.  For advection in the radial direction,
$\varpi^*$ must be substituted for $\varpi$ in (\ref{smooth}) so that
the radius is evaluated at the same position as the interpolated
quantities; see equation~(\ref{varpi-star}).

\subsection{Time Step Controls}\label{sec-courant}

The finite difference equations that have been developed are used to
evolve a numerical model of a physical system forward in time by
taking discrete time steps.  The time steps cannot be arbitrarily
large, as the system of equations can become unstable (Bowers \&
Wilson 1991).  We now discuss the restrictions placed on the time step
to produce accurate and stable solutions.

Whenever a quantity is being transported (or a signal is being
propagated) through a grid, it is important that the time step be
small enough so that the quantity or signal cannot cross an entire
grid zone in a single time step.  This is the well known Courant
condition (Courant, Friedrichs, \& Lewy 1928) given by $v\; \Delta t /
\Delta x \leq 1$ in 1-D. It leads to two explicit criteria that the
time step in our code must satisfy.

In an Eulerian code the fluid itself is transported through the grid,
with a velocity $\vec{v}_r=\vec{v}-\vec{v}_g$ relative to the grid.
The Courant condition for fluid transport applied in each of the three
coordinate directions is thus
\pagebreak[1]{\samepage\begin{eqnarray}
\Delta t \leq (\Delta t_\varpi^T)_{j+\frac{1}{2},k+\frac{1}{2},l+\frac{1}{2}}&
\equiv &
\frac{\Delta \varpi_{j+\frac{1}{2}}}
{\left|v_{j+\frac{1}{2},k+\frac{1}{2},l+\frac{1}{2}}-{v_g}_{j+\frac{1}{2}}\right|},
\nonumber\\
\Delta t \leq (\Delta t_z^T)_{j+\frac{1}{2},k+\frac{1}{2},l+\frac{1}{2}}&
\equiv &
\frac{\Delta
z_{k+\frac{1}{2}}}{\left|u_{j+\frac{1}{2},k+\frac{1}{2},l+\frac{1}{2}}-{u_g}_{k+\frac{1}{2}}
\right|},
\label{ttrans}\\
\Delta t \leq (\Delta t_\varphi^T)_{j+\frac{1}{2},k+\frac{1}{2},l+\frac{1}{2}}&
\equiv &
\frac{\varpi\Delta \varphi_{l+\frac{1}{2}}}
{\left|w_{j+\frac{1}{2},k+\frac{1}{2},l+\frac{1}{2}}-{w_g}_{l+\frac{1}{2}}\right|}.  \nonumber
\end{eqnarray}}\pagebreak[1]Note that the time step constraint varies for each
zone and for each coordinate
direction, so that (\ref{ttrans}) actually specifies $3N_{\rm zones}$
time constraints, where $N_{\rm zones}$ is the total number of grid
zones.  Since all of these contraints are satisfied if we satisfy the
most restrictive of them, we define the transport time step condition
to be
\begin{equation}
\Delta t\leq\Delta t^T,
\end{equation}
where
\begin{equation}
\Delta t^T\equiv\min\left[(\Delta
t_\varpi^T)_{j+\frac{1}{2},k+\frac{1}{2},l+\frac{1}{2}},
(\Delta t_z^T)_{j+\frac{1}{2},k+\frac{1}{2},l+\frac{1}{2}}, (\Delta
t_\varphi^T)_{j+\frac{1}{2},k+\frac{1}{2},l+\frac{1}{2}}\right]
\label{dttrans}
\end{equation}
and the minimum is taken over the entire grid.

Although sound waves do not physically transport fluid through the
grid, they do propagate changes in physical variables such as density
and pressure.  Therefore, the Courant condition must be satified for
the propagation of sound waves through the grid.  The adiabatic sound
speed in the fluid is
\begin{equation}
c_s=\sqrt{\frac{\partial p}{\partial\rho}},
\end{equation}
which, for the ideal gas EOS (\ref{ideal-gas}), is simply
\begin{equation}
c_s=\sqrt{(\gamma-1)\varepsilon}.
\end{equation}
This is the velocity of the sound wave {\it with respect to the
fluid\/} through which it travels.  If the fluid itself is moving
through the grid, then the speed of sound with respect to the grid
varies with the direction of propagation, but has a maximum of
$|\vec{v}_r|+c_s$.  Thus the Courant condition for sound wave
propagation in each of the three coordinate directions is
\pagebreak[1]{\samepage\begin{eqnarray}
\Delta t \leq (\Delta t_\varpi^S)_{j+\frac{1}{2},k+\frac{1}{2},l+\frac{1}{2}}&
\equiv &
\frac{\Delta \varpi_{j+\frac{1}{2}}}
{\left|v_{j+\frac{1}{2},k+\frac{1}{2},l+\frac{1}{2}}-{v_g}_{j+\frac{1}{2}}\right|+c_s},
\nonumber\\
\Delta t \leq (\Delta t_z^S)_{j+\frac{1}{2},k+\frac{1}{2},l+\frac{1}{2}}&
\equiv &
\frac{\Delta
z_{k+\frac{1}{2}}}{\left|u_{j+\frac{1}{2},k+\frac{1}{2},l+\frac{1}{2}}-{u_g}_{k+\frac{1}{2}}
\right|+c_s},
\label{tcour}\\
\Delta t \leq (\Delta t_\varphi^S)_{j+\frac{1}{2},k+\frac{1}{2},l+\frac{1}{2}}&
\equiv &
\frac{\varpi\Delta \varphi_{l+\frac{1}{2}}}
{\left|w_{j+\frac{1}{2},k+\frac{1}{2},l+\frac{1}{2}}-{w_g}_{l+\frac{1}{2}}\right|+c_s}
{}.
\nonumber
\end{eqnarray}}\pagebreak[1]Again, the time step is limited by the
most restrictive of these conditions, so the sound speed Courant
condition becomes
\begin{equation}
\Delta t\leq\Delta t^S,
\label{concour}
\end{equation}
where
\begin{equation}
\Delta t^S\equiv\min\left[(\Delta
t_\varpi^S)_{j+\frac{1}{2},k+\frac{1}{2},l+\frac{1}{2}},
(\Delta t_z^S)_{j+\frac{1}{2},k+\frac{1}{2},l+\frac{1}{2}}, (\Delta
t_\varphi^S)_{j+\frac{1}{2},k+\frac{1}{2},l+\frac{1}{2}}\right].
\label{dtcour}
\end{equation}

The presence of artificial viscosity introduces an explicit diffusion
into the system and therefore leads to an additional restriction on
the time step.  Considering only the compressional viscosity, the
condition for stability in each direction is (Bowers \& Wilson 1991)
\begin{equation}
\Delta t\leq\frac{1}{4}\frac{\Delta x_a}{\sqrt{Q^{aa}}}
\;\;\;\;\;\mbox{(no summation)}.
\end{equation}
Satisfying this for all three coordinate directions in each grid zone
again yields a set of $3N_{\rm zones}$ constraint conditions, of which
we must satisfy the most restrictive.  Using expressions
(\ref{Qrr})--(\ref{Qtt}) for the artificial viscosity tensor we define
\pagebreak[1]{\samepage\begin{eqnarray}
(\Delta t_\varpi^Q)_{j+\frac{1}{2},k+\frac{1}{2},l+\frac{1}{2}}&
\equiv &
\frac{\Delta \varpi_{j+\frac{1}{2}}}
{4\sqrt{C_Q}\left|v_{j+1,k+\frac{1}{2},l+\frac{1}{2}}-v_{j,k+\frac{1}{2},l+\frac{1}{2}}\right|},
\nonumber\\
(\Delta t_z^Q)_{j+\frac{1}{2},k+\frac{1}{2},l+\frac{1}{2}}& \equiv &
\frac{\Delta z_{k+\frac{1}{2}}}
{4\sqrt{C_Q}\left|u_{j+\frac{1}{2},k+1,l+\frac{1}{2}}-u_{j+\frac{1}{2},k,l+\frac{1}{2}}\right|},
\label{tshock}\\
(\Delta t_\varphi^Q)_{j+\frac{1}{2},k+\frac{1}{2},l+\frac{1}{2}}&
\equiv &
\frac{\varpi\Delta \varphi_{l+\frac{1}{2}}}
{4\sqrt{C_Q}\left|w_{j+\frac{1}{2},k+\frac{1}{2},l+1}-w_{j+\frac{1}{2},k+\frac{1}{2},l}\right|}
{}.
\nonumber
\end{eqnarray}}\pagebreak[1]Then
the artificial viscosity constraint becomes
\begin{equation}
\Delta t \leq \Delta t^Q,
\label{conshock}
\end{equation}
where
\begin{equation}
\Delta t^Q\equiv\min\left[(\Delta
t_\varpi^Q)_{j+\frac{1}{2},k+\frac{1}{2},l+\frac{1}{2}},
(\Delta t_z^Q)_{j+\frac{1}{2},k+\frac{1}{2},l+\frac{1}{2}}, (\Delta
t_\varphi^Q)_{j+\frac{1}{2},k+\frac{1}{2},l+\frac{1}{2}}\right] .
\label{dtshock}
\end{equation}

The time step must be less than the three values defined by
(\ref{dttrans}), (\ref{dtcour}), and (\ref{dtshock}). Note that the
sound speed condition is always more restrictive than the transport
condition since $\Delta t^S\leq\Delta t^T$, where the equality only
holds in the case where pressure is independent of density and the
sound speed $c_s = 0$.  We therefore ignore the transport constraint
and concentrate only on the sound speed and artificial viscosity
constraints (\ref{concour}) and (\ref{conshock}).

These time step restrictions are the minimal constraint conditions
based on a linearized stability analysis (Bowers \& Wilson 1991;
Press, et al. 1992).  To allow for nonlinear effects we therefore
introduce two constant parameters, $C^S< 1$ and $C^Q< 1$, which are
given as input to the program, and impose the condition
\begin{equation}
\Delta t \leq \min\left[C^S\Delta t^S,C^Q\Delta t^Q\right].
\end{equation}
The values chosen for the constants $C^S$ and $C^Q$ are important.  If
they are too large, the accuracy of the simulation may be compromised;
if they are too small, the number of cycles, and hence the CPU time,
required for the calculation is increased. We usually run with $C^S
\approx C^Q \approx 0.3$.

There is one final condition that should be applied.  The time step is
recalculated for each cycle, and therefore is constantly varying.  If
the time step changes by too much from one cycle to the next,
irregularities can arise in the computation.  To avoid this, we limit
the maximum change from one time step to the next by introducing
another constant, $C^\delta>1$, and requiring
\begin{equation}
\Delta t^{n+\frac{1}{2}} = \min\left[C^S\Delta t^S,C^Q\Delta t^Q,
C^\delta t^{n-\frac{1}{2}}\right].
\label{tnhalf}
\end{equation}
It is usually sufficient to choose $C^\delta$ such that the time step
cannot grow by more than 10\%--20\% per cycle.

The time step calculated in (\ref{tnhalf}) is used to update zone
centered quantities, defined at integral time levels $t^n$.  Face
centered quantities are defined at half-integral time levels
$t^{n-\frac{1}{2}}$, and these are updated using the time step
\begin{equation}
\Delta t^n = \textstyle{\frac{1}{2}}\left(\Delta t^{n-\frac{1}{2}} + \Delta
t^{n+\frac{1}{2}} \right).
\label{tn}
\end{equation}
When $\Delta t^{n+\frac{1}{2}}$ and $\Delta t^n$ are calculated, the
scalar quantities are known at time level $n$ and the vector
quantities are known at time level $n-\frac{1}{2}$, and it is these
values that are used in equations (\ref{tcour}) and (\ref{tshock}).
Using conservative values for the constants in (\ref{tnhalf}) should
guarantee that the constraints will be satisfied for the values of the
variables that are coincident with the time step even though they have
been calculated using values that are up to a full time step behind.

\section{NEWTONIAN GRAVITATIONAL FIELD:  POISSON'S EQUATION}\label{sec-pois}

The Newtonian gravitational potential $\Phi$ is obtained by solving
Poisson's equation (\ref{poisson-cyl}).  Expressed in finite
difference form this gives\pagebreak[1] {\samepage
\begin{eqnarray}
\left[\varpi_{j+1}\frac{\Phi_{j+\frac{3}{2}}-\Phi_{j+\frac{1}{2}}}{\Delta\varpi_{j+1}}
-\varpi_{j}\frac{\Phi_{j+\frac{1}{2}}-\Phi_{j-\frac{1}{2}}}{\Delta\varpi_{j}}
\right]_{l+\frac{1}{2},k+\frac{1}{2}}\Delta
z_{k+\frac{1}{2}}\Delta\varphi_{l+\frac{1}{2}}&&\nonumber\\
+\left[\frac{\Phi_{k+\frac{3}{2}}-\Phi_{k+\frac{1}{2}}}{\Delta
z_{k+1}} -\frac{\Phi_{k+\frac{1}{2}}-\Phi_{k-\frac{1}{2}}}{\Delta
z_{k}}
\right]_{j+\frac{1}{2},l+\frac{1}{2}}(\varpi\Delta\varpi)_{j+\frac{1}{2}}\Delta\varphi_{l+\frac{1}{2}}
&&\nonumber\\
+\left[\frac{\Phi_{l+\frac{3}{2}}-\Phi_{l+\frac{1}{2}}}{\Delta\varphi_{l+1}}
-\frac{\Phi_{l+\frac{1}{2}}-\Phi_{l-\frac{1}{2}}}{\Delta\varphi_{l}}
\right]_{j+\frac{1}{2},k+\frac{1}{2}}\frac{\Delta\varpi_{j+\frac{1}{2}}}{\varpi_{j+\frac{1}{2}}}
\Delta z_{k+\frac{1}{2}}\;&&\nonumber\\
-4\pi G \rho_{j+\frac{1}{2},k+\frac{1}{2},l+\frac{1}{2}}\Delta
V_{j+\frac{1}{2},k+\frac{1}{2},l+\frac{1}{2}}&=&0.
\;\;\;\;\;\;\;\;\;\end{eqnarray}}\pagebreak[1]
Rearranging terms, we find\pagebreak[1] {\samepage
\begin{eqnarray}
f_1\Phi_{j+\frac{3}{2},k+\frac{1}{2},l+\frac{1}{2}} + f_2
\Phi_{j-\frac{1}{2},k+\frac{1}{2},l+\frac{1}{2}} +
f_3\lefteqn{\Phi_{j+\frac{1}{2},k+\frac{1}{2},l+\frac{3}{2}}}
&&\nonumber\\ + f_4 \Phi_{j+\frac{1}{2},k+\frac{1}{2},l-\frac{1}{2}} +
f_5\Phi_{j+\frac{1}{2},k+\frac{3}{2},l+\frac{1}{2}} &&\nonumber\\ +
f_6 \Phi_{j+\frac{1}{2},k-\frac{1}{2},l+\frac{1}{2}} +
f_7\Phi_{j+\frac{1}{2},k+\frac{1}{2},l+\frac{1}{2}} &=& 4\pi G
\rho_{j+\frac{1}{2},k+\frac{1}{2},l+\frac{1}{2}}\Delta
V_{j+\frac{1}{2},k+\frac{1}{2},l+\frac{1}{2}},
\label{pois-grid}\;\;\;\;\;\;\;
\end{eqnarray}}\pagebreak[1]
where
\begin{eqnarray}
f_1=\frac{\varpi_{j+1}\Delta z_{k+\frac{1}{2}}
\Delta\varphi_{l+\frac{1}{2}}}{\Delta\varpi_{j+1}}\;\;
&,& f_2=\frac{\varpi_{j}\Delta z_{k+\frac{1}{2}}
\Delta\varphi_{l+\frac{1}{2}}}{\Delta\varpi_{j}}
\;\;\;\;\:\;,  \nonumber \\
f_3=\frac{\Delta\varpi_{j+\frac{1}{2}}\Delta
z_{k+\frac{1}{2}}}{\varpi_{j+\frac{1}{2}}\Delta\varphi_{l+1}}\;\;\;\;\;\;\;\;\:
&,& f_4=\frac{\Delta\varpi_{j+\frac{1}{2}}\Delta
z_{k+\frac{1}{2}}}{\varpi_{j+\frac{1}{2}}\Delta\varphi_{l}}
\;\;\;\;\;\;\;\;\:,  \nonumber \\
f_5=\frac{\varpi_{j+\frac{1}{2}}\Delta\varpi_{j+\frac{1}{2}}\Delta\varphi_{l+\frac{1}{2}}}{\Delta
z_{k+1}} &,&
f_6=\frac{\varpi_{j+\frac{1}{2}}\Delta\varpi_{j+\frac{1}{2}}\Delta\varphi_{l+\frac{1}{2}}}{\Delta
z_{k}}\;,
\label{coeff}
\end{eqnarray}
and
\begin{displaymath}
f_7=-(f_1+f_2+f_3+f_4+f_5+f_6).
\end{displaymath}
We can rewrite equation~(\ref{pois-grid}) in matrix form by relabeling
the 3-D arrays $\rho_{j+\frac{1}{2},k+\frac{1}{2},l+\frac{1}{2}}$ and
$\Phi_{j+\frac{1}{2},k+\frac{1}{2},l+\frac{1}{2}}$ so that they are in
1-D form $\hat{\rho}_n$ and $\hat{\Phi}_n$.  For a grid with
dimensions $N_{\rm zones} = N_{\varpi}\times N_z\times N_{\varphi}$,
the 1-D index $N$ can be written
\begin{equation}
N=L+(J-1)N_{\varphi}+(K-1)N_{\varphi}N_{\varpi}.
\label{1-D-index}
\end{equation}
where we have defined $J \equiv j + \frac{1}{2}$, $K \equiv k +
\frac{1}{2}$, and $L \equiv l + \frac{1}{2}$ for convenience.
With this,~(\ref{pois-grid}) becomes
\begin{eqnarray}
f_1 \hat\Phi_{N+N_{\varphi}} + f_2 \hat\Phi_{N - N_{\varphi}} + f_3
\hat\Phi_{N+1} + f_4 \hat\Phi_{n-1} & + & \nonumber \\ f_5
\hat\Phi_{N+ N{\varphi}N_{\varpi}} + f_6 \hat\Phi_{N-
N{\varphi}N_{\varpi}} + f_7 \hat\Phi_N & = & 4 \pi G (\hat\rho \Delta
V)_N .
\label{pois-a}
\end{eqnarray}
By inspection we see that~(\ref{pois-a}) is in matrix form
\begin{equation}
A_{MN} \hat\Phi_N = 4 \pi G(\hat\rho \Delta V)_M,
\label{pois-mat}
\end{equation}
where $A_{MN}$ is a sparse, banded matrix that is tridiagonal with 2
additional diagonal bands above and below the main diagonal; such a
matrix is called ``tridiagonal with fringes'' (Press, et al. 1992),

We have used the preconditioned conjugate gradient (PCG) method
(Press, et al. 1992) to solve Poisson's equation in the
form~(\ref{pois-mat}).  We have found that this produces accurate
and stable solutions, as will be demonstrated by the test cases
reported later in this section and in \S~\ref{tests}.  The PCG
method was derived to solve a large, sparse system of equations
in which the matrix $A_{MN}$ is symmetric and positive
definite, and we have found it to perform well in cases where the
matrix is nearly symmetric.  From equations~(\ref{pois-a})
and~(\ref{coeff}) we see that $A_{MN}$ is exactly symmetric when a
uniform grid is used, and is nearly symmetric (typically within
a few percent) for nonuniform zoning with zoning ratios near
unity.  These conditions are always met in our simulations.

The PCG method is an iterative technique that works as follows (Clancy
1989; Press, et al. 1992).  The equation to be solved is written in
the form
\begin{equation}
A \cdot \vec x = \vec b,
\label{mat-eq}
\end{equation}
where $A$ is a matrix and the dot indicates matrix multiplication.
Let
\begin{equation}
\vec r\,^i = \vec b - A \cdot \vec x\,^i
\label{resid}
\end{equation}
be the residual, and $\vec x\,^i$ be the solution on the $i$th
iteration.  Next, form
\begin{equation}
\vec z\,^i = \tilde A^{-1} \cdot \vec r\,^i,
\label{zvec}
\end{equation}
where the matrix $\tilde A$ is called the preconditioner.  Then,
calculate
\begin{equation}
\beta^i = \frac{\vec z\,^i \cdot \vec r\,^i}
{\vec z\,^{i-1} \cdot \vec r\,^{i-1}},
\label{beta}
\end{equation}
\begin{equation}
\vec p\,^i = \vec z\,^i + \beta^i \vec p\,^{i-1},
\label{pvec}
\end{equation}
and
\begin{equation}
\alpha^i = \frac{\vec z\,^i \cdot \vec r\,^i}
{\vec p\,^i \cdot (A \cdot \vec p\,^i)} .
\label{alpha}
\end{equation}
The improved estimate to the solution is
\begin{equation}
\vec x\,^{i+1} = \vec x\,^i + \alpha^i \vec p\,^i,
\label{new-x}
\end{equation}
and the new residual is
\begin{equation}
\vec r\,^{i+1} = \vec r\,^i + \alpha(A \cdot \vec p\,^i).
\label{new-resid}
\end{equation}
To start the process, choose $\vec x\,^0$ to be an initial guess for
the solution and use~(\ref{resid}) and~(\ref{zvec}) to form $\vec
r\,^0$ and $\vec z\,^0$, respectively.  Let $\beta^0 = 0$, and then
proceed to iterate beginning with equation~(\ref{pvec}).  Note that
equation~(\ref{new-x}) guarantees that $r\,^{i+1}$
in~(\ref{new-resid}) is in fact the residual given by~(\ref{resid}).

We continue iterating until we have converged to the solution for
$\Phi$ to within a specified tolerance $tol$.  We monitor 2
quantities,
\begin{equation}
\epsilon_1 = \frac{1}{N_{grid}}
\left( \sum_{grid} \left[ 1 - \frac{\Phi^{i+1}}{\Phi^i} \right ] ^2
\right ) ^{1/2}
\label{eps1}
\end{equation}
and
\begin{equation}
\epsilon_2 = \frac{\sum_{grid} (\nabla^2 \Phi - 4 \pi G \rho)^2}
{\sum_{grid} (4 \pi G \rho)^2},
\label{eps2}
\end{equation}
and stop iterating when either $\epsilon_1 < tol$ or $\epsilon_2 <
tol$.  To achieve accuracy of $\sim 1 \%$ in $\nabla \Phi$, we need
$tol \lesssim 10^{-5}$.

There are various choices for the preconditioning matrix $\tilde A$.
We have chosen to let $\tilde A$ be the diagonal part of $A$,
\begin{equation}
\tilde A = {\rm diag}(A),
\label{diag-scale}
\end{equation}
to minimize the amount of storage required.  This is known as diagonal
scaling (Meijerink \& Van Der Vorst 1981; Press, et al. 1992).  We
have implemented this method using only 3 additional 3-D arrays, one
for each of the vectors $\vec r$, $\vec z$, and $\vec p$; the vector
$A \cdot \vec p$ in equation~(\ref{alpha}) can also be stored in the
array used to hold $\vec z$.  Two of these arrays are shared storage
spaces with the temporary arrays needed to solve the hydrodynamical
equations, and so solving Poisson's equation requires only one
additional 3-D array.  Using the indexing
convention~(\ref{1-D-index}), the coefficients~(\ref{coeff}) are
calculated for each $N$ within the loop to form the components of the
matrices $A$ and $\tilde A$ as needed.  Overall, this method yields a
significant reduction in memory requirements over related methods.

We ran comparison tests of our Poisson solver with the routine SITRSOL
from the Cray library of scientific subroutines LIBSCI (Cray Research
1992).  This routine is a generic solver for real, sparse linear
systems of the form~(\ref{mat-eq}), allowing for a variety of solution
methods and preconditioning options.  We modeled uniform density dust
ellipsoids with eccentricities up to 0.993 and compared the
numerically calculated values of $\Phi$ with the analytic expressions
given by Chandrasekhar (1969); cf. \S~\ref{sec-ellip-tests} below.  We
ran with $32 \times 32 \times 32$ and $64 \times 32 \times 64\;
(\varpi,z,\varphi)$ grids, with zoning ratios in $\varpi$ and $z$
varying from 1.0 (uniformly zoned) to 1.05.  The outer boundary of the
grid was generally set at a distance of twice the largest extent of
the mass distribution to produce good boundary conditions; see below.
In terms of accuracy, all of the SITRSOL algorithms that converged
produced results that matched the solution given by our Poisson
solver.  In all cases the SITRSOL algorithms used the full matrix
$A_{MN}$, without assuming that it is symmetric.  Since they gave the
same result as our routine, this is confirmation that the our Poisson
solver is reaching the right solution even with the slight
non-symmetry in $A_{MN}$ due to nonuniform zoning.  When evaluated for
efficiency, both in terms of memory and computational speed, our
solver was clearly superior to the other options we investigated.
Although the Cray routine is fully optimized, it required
approximately 10 times more CPU time, regardless of the choice of
method or the mass distribution.  Furthermore, it increased the memory
requirements of the code by a factor of 10--30, depending on the
choice of method and preconditioner.  These differences are due at
least in part to the fact that our routine was designed for the
specific problem we are solving, never requires the storage of the
entire matrix being inverted, and takes advantage of existing
temporary arrays.  Since the Cray solver is a generic sparse matrix
routine it cannot efficiently take advantage of what is known for the
specific case at hand.

The solution of Poisson's equation also requires specifying boundary
conditions for the potential.  These are given at the edges of the
grid using a spherical multipole expansion
\begin{equation}
\Phi(r,\theta,\varphi) =
\sum_{n=0}^{\infty}\sum_{m=-n}^{n}\frac{4\pi}{2n+1}q_{nm}
\frac{Y_{nm}(\theta ,\varphi)}{r^{n+1}},
\label{bcgrav}
\end{equation}
where
\begin{equation}
q_{nm}=\int_V Y^{*}_{nm}r^n\rho d^3V.
\label{mass-moment}
\end{equation}
Here $r,\theta,\varphi$ are the standard spherical coordinates, the
asterisk indicates the complex conjugate, and the integral is taken
over the entire source.  We carry the expansion out to order $n=8$.

To calculate these boundary conditions numerically, we use a set of
real harmonic functions $\tilde{y}_{nm}$ that are related to the
standard spherical harmonics by the equation
\begin{displaymath}
Y_{nm}(\theta,\varphi)=\sqrt{\frac{2n+1}{4\pi}}\tilde{y}_{nm}(\theta)e^{im\varphi}.
\end{displaymath}
The $\tilde y_{nm}$ are simply the associated Legendre functions with
a non-standard normalization,
\begin{displaymath}
\tilde{y}_{nm}(\theta) = (-1)^m\sqrt{\frac{(n-m)!}{(n+m)!}}P_{nm}(\cos\theta).
\end{displaymath}
Thus, they obey the recursion relations
\begin{displaymath}
\tilde{y}_{nm}(\theta)=\frac{2n-1}{\sqrt{(n-m)(n+m)}}\cos(\theta)
\tilde{y}_{n-1,m}-\sqrt{\frac{(n-m-1)(n+m-1)}{(n-m)(n+m)}}\tilde{y}_{n-2,m}
\end{displaymath}
and
\begin{displaymath}
\tilde{y}_{nm}(\theta)=\frac{2(m-1)}{\sqrt{(n+m)(n-m+1)}}\cot(\theta)
\tilde{y}_{n,m-1}
-\sqrt{\frac{(n-m+2)(n+m-1)}{(n+m)(n-m+1)}}\tilde{y}_{n,m-2},
\end{displaymath}
with
\begin{displaymath}
\tilde{y}_{00}(\theta)=1,
\end{displaymath}
\begin{displaymath}
\tilde{y}_{10}(\theta)=\cos\theta,
\end{displaymath}
and
\begin{displaymath}
\tilde{y}_{11}(\theta)=-\frac{\sin\theta}{\sqrt{2}}.
\end{displaymath}
With this, the integral (\ref{mass-moment}) can be written in finite
difference form as
\begin{equation}
\tilde{q}_{nm}=\sum_{j,l,k}r_{jk}^n\tilde{y}_{nm}\left(\theta_{jk}\right)e^{-im\varphi_l}
(\rho\Delta V)_{j+\frac{1}{2},k+\frac{1}{2},l+\frac{1}{2}},
\end{equation}
where $r_{jk}=\sqrt{\varpi_{j+\frac{1}{2}}^2+z_{k+\frac{1}{2}}^2}$,
$\theta_{jk}=\tan^{-1}\frac{z_{k+\frac{1}{2}}}{\varpi_{j+\frac{1}{2}}}$,
and the symbol $\tilde q_{nm}$ indicates that we have used $\tilde
y_{nm}$ instead of $Y_{nm}$ in~(\ref{mass-moment}).  Then
(\ref{bcgrav}) becomes
\begin{equation}
\Phi_{j+\frac{1}{2},k+\frac{1}{2},l+\frac{1}{2}}
=\sum_{n=0}^{n_{\hbox{\tiny{max}}}}\left\{
\tilde{q}_{n0}\frac{\tilde{y}_{n0}(\theta_{jk})}{r_{jk}^{n+1}} +
\sum_{m=1}^{n}
2\Re\left[\tilde{q}_{nm}e^{im\varphi_{l+\frac{1}{2}}}\right]
\frac{\tilde{y}_{nm}(\theta_{jk})}{r_{jk}^{n+1}}\right\},
\label{phi-bc}
\end{equation}
where $n_{\hbox{\scriptsize{max}}}$ indicates the highest order
multipole, currently 8 in our case, and $\Re$ is used to indicate that
we take the real part of the quantity in square brackets.  Multipole
moments $\tilde{q}_{nm}$ with $n+m$ odd are identically zero due to
the reflection symmetry through $z=0$.

Note that the boundary conditions~(\ref{phi-bc}) are determined by the
source and are calculated in a separate subroutine prior to solving
Poisson's equation.

It is a simple matter to increase the number of terms in the expansion
of the potential~(\ref{phi-bc}) at the boundaries of the grid.
However, due to the discrete nature of the grid, we must not take the
expansion out too far.  This decomposition of the potential into
multipoles is based on the orthogonality of the spherical harmonics,
which must be preserved on the grid.  Thus, the accuracy of the
calculation of the boundary conditions depends on accurately resolving
the functions used in the expansion.

For example, consider the function $\cos m\varphi$, where $m$ is equal
to the number of angular grid zones $N_{\varphi}$.  If the angular
grid is equally spaced, then this function will yield a value of $1$
at each point on the grid since the argument will always be a multiple
of $2\pi$.  For an axisymmetric distribution, numerically integrating
this function will give nonzero values for multipole moments with
$m=N_{\varphi}$, although all multipole moments with $m\neq 0$ should
be zero by symmetry.  Therefore, we must have the order of the
multipole expansion $n_{\hbox{\scriptsize{max}}} < N_{\varphi}$ at
least; in general we keep $n_{\hbox{\scriptsize{max}}} <
N_{\varphi}/4$.

We expect a similar restriction on $n_{\hbox{\scriptsize{max}}}$ from
the evaluation of the Legendre polynomials at discrete grid locations.
Estimating this restriction is more difficult for two reasons.  First,
the Legendre polynomials themselves are more complicated then the
simple trigonometric functions.  Also, since the argument of the
polynomials, $\cos\theta \equiv z/\varpi$, is not a natural variable
on our grid, the sampling distribution for a discrete realization of
the functions is a complicated function of the $z$ and $\varpi$ grid
spacings, with some regions sampled more densely than others.

We must thus take care to limit the number of terms in the multipole
expansion~(\ref{phi-bc}).  We have found that errors in the
calculation of the boundary values for $\Phi$ are most likely to occur
near the $z$-axis.  For example, for a uniform density oblate spheroid
of eccentricity 0.995 on a $64\times 32\times 32$ ($\varpi,z,\varphi$)
grid with the grid boundary set at twice the equatorial radius, the
multipole expansion~(\ref{phi-bc}) gave values for $\Phi$ that matched
the analytic values to within $1\%$ over the entire boundary except
for the zones on the $z$-axis.  On the $z$-axis there were large
errors ($\sim 100\%$) when the expansion was taken to
$n_{\hbox{\scriptsize{max}}}=10$.  These did not improve when
additional terms in the expansion were included, but were reduced to
$\sim 1\%$ when the order of the expansion was reduced to
$n_{\hbox{\scriptsize{max}}}=8$.

For a given angular resolution $N_{\varphi}$, the accuracy of the
multipole expansion can be improved only by moving the boundary
farther from the surface of the mass distribution to reduce the
magnitude of the higher order terms that have been discarded.  We have
found that for significantly aspherical configurations it is best to
set both the $\varpi$ and $z$ boundaries at least a factor of 2 larger
than the largest dimension of the mass distribution.  For example,
when this condition was imposed for a uniform density triaxial
ellipsoid with axis ratios 4:2:1 (in the $x$, $y$, and $z$-directions,
respectively), the potential deviated from the expected value by $ <
1\%$ throughout the grid on a $64\times 32\times 64$
($\varpi,z,\varphi$) uniform grid with
$n_{\hbox{\scriptsize{max}}}=8$.  When the ratio of the grid boundary
to the largest axis was reduced from 2 to $1.5$, maximum errors of
several percent were encountered for the same parameters.

\section{EXTRACTION OF GRAVITATIONAL RADIATION}\label{extract}

\subsection{Quadrupole Formulae}\label{sec-grdefs}

We calculate the gravitational radiation produced in these models
using the quadrupole approximation (Misner, Thorne and Wheeler 1973).
The spacetime metric is
\begin{equation}
	g_{\mu\nu} = \eta_{\mu\nu} + h_{\mu\nu} ,
\label{metric}
\end{equation}
where $\mu,\nu = 0,1,2,3$, $\eta_{\mu\nu} = {\rm diag}(-1,1,1,1)$ is
the metric of flat spacetime, and $|h_{\mu\nu}| \ll 1$.  The
gravitational wave amplitude at distance $r$ from the source is given
by the transverse-traceless (TT) components of the metric perturbation
$h_{ij}$,
\begin{equation}
h^{TT}_{ab} = \frac{2}{r}\frac{G}{c^4} {\skew6\ddot{
{I\mkern-6.8mu\raise0.3ex\hbox{-}}}}^{\,TT}_{ab},
\label{g-amp}
\end{equation}
where the reduced mass quadrupole moment tensor in Cartesian
coordinates is
\begin{equation}
{I\mkern-6.8mu\raise0.3ex\hbox{-}}_{ab}(t) = \int
\rho\left(\vec{x},t\right)\left(x_ax_b-\textstyle{\frac{1}{3}}\delta_{ab}r^2\right)dV.
\label{ibar}
\end{equation}
Here, spatial indices $a,b = 1,2,3$, the dot indicates a time
derivative $d/dt$, and the integral is performed over the volume of
the source.  The TT part of the quadrupole moment is defined by
\begin{equation}
 {\skew6\ddot{ {I\mkern-6.8mu\raise0.3ex\hbox{-}}}}^{\,TT}_{ab} =
P_{ac} {\skew6\ddot{ {I\mkern-6.8mu\raise0.3ex\hbox{-}}}}_{cd}P_{db} -
\textstyle{\frac{1}{2}} P_{ab}P_{cd} {\skew6\ddot{
{I\mkern-6.8mu\raise0.3ex\hbox{-}}}}_{cd},
\label{TT-proj}
\end{equation}
where the projection operator $P_{ab} = \delta_{ab}-n_an_b$ removes
the parts of the quadrupole tensor in the propagation direction and
$n_a=x_a/r$ are the components of the radial unit vector from the
source to the field point.  The gravitational wave luminosity of the
source is
\begin{equation}
L = {\frac{1}{5}} \left\langle
{I\mkern-6.8mu\raise0.3ex\hbox{-}}^{(3)}_{ab}
{I\mkern-6.8mu\raise0.3ex\hbox{-}}^{(3)}_{ab} \right \rangle,
\label{g-lum}
\end{equation}
where the superscript $(3)$ indicates 3 time derivatives, there is an
implied sum on repeated indices, and the angle brackets are used to
denote an average over several wavelengths.  The energy emitted as
gravitational radiation is
\begin{equation}
\Delta E = \int L \; dt.
\label{delta-E}
\end{equation}
Finally, the angular momentum carried away by the gravitational waves
is
\begin{equation}
\frac{dJ_a}{dt} = {\frac{2}{5}} \epsilon_{abd}
 \left\langle {I\mkern-6.8mu\raise0.3ex\hbox{-}}^{(2)}_{bf}
{I\mkern-6.8mu\raise0.3ex\hbox{-}}^{(3)}_{fd} \right \rangle,
\label{j-lum}
\end{equation}
where $\epsilon_{abd}$ is the totally antisymmetric Levi-Civita
tensor.

If we use an orthonormal spherical coordinate system defined by the
unit vectors
\begin{displaymath}
\widehat{e}_r\equiv\frac{\partial}{\partial r},\;\;\;\;\;
\widehat{e}_\theta\equiv\frac{1}{r}\frac{\partial}{\partial\theta},\;\;\;\;\;
\widehat{e}_\phi\equiv\frac{1}{r\sin\theta}\frac{\partial}{\partial\phi},
\end{displaymath}
with the center of mass of the source located at the origin, then the
TT part of ${I\mkern-6.8mu\raise0.3ex\hbox{-}}_{ab}$ has only four
non-vanishing components.  Expressed in terms of the Cartesian
components~(\ref{ibar}) these are (Kochanek, et al. 1990)\pagebreak[1]
{\samepage
\begin{eqnarray}
{I\mkern-6.8mu\raise0.3ex\hbox{-}}_{\theta\theta}&=&
({I\mkern-6.8mu\raise0.3ex\hbox{-}}_{xx}\cos^2\phi+
{I\mkern-6.8mu\raise0.3ex\hbox{-}}_{yy}\sin^2\phi+
{I\mkern-6.8mu\raise0.3ex\hbox{-}}_{xy}\sin 2\phi)\cos^2\theta
\nonumber\\&&\;\;\;\;\;\;\;\;\;\;\;\;\;\;\;+
{I\mkern-6.8mu\raise0.3ex\hbox{-}}_{zz}\sin^2\theta -
({I\mkern-6.8mu\raise0.3ex\hbox{-}}_{xz}\cos\phi+
{I\mkern-6.8mu\raise0.3ex\hbox{-}}_{yz}\sin\phi)\sin 2\theta,
\nonumber\\ {I\mkern-6.8mu\raise0.3ex\hbox{-}}_{\phi\phi}&=&
{I\mkern-6.8mu\raise0.3ex\hbox{-}}_{xx}\sin^2\phi+
{I\mkern-6.8mu\raise0.3ex\hbox{-}}_{yy}\cos^2\phi
-{I\mkern-6.8mu\raise0.3ex\hbox{-}}_{xy}\sin 2\phi, \label{Ispher}\\
{I\mkern-6.8mu\raise0.3ex\hbox{-}}_{\theta\phi}&=&
{I\mkern-6.8mu\raise0.3ex\hbox{-}}_{\phi\theta} \nonumber\\ &=&
\textstyle{\frac{1}{2}}({I\mkern-6.8mu\raise0.3ex\hbox{-}}_{yy}-
{I\mkern-6.8mu\raise0.3ex\hbox{-}}_{xx})\cos\theta\sin
2\phi\nonumber\\&&
\;\;\;\;\;\;\;\;\;\;\;\;\;\;\;+
{I\mkern-6.8mu\raise0.3ex\hbox{-}}_{xy}\cos\theta\cos 2\phi +
({I\mkern-6.8mu\raise0.3ex\hbox{-}}_{xz}\sin\phi -
{I\mkern-6.8mu\raise0.3ex\hbox{-}}_{yz}\cos\phi)\sin\theta.\nonumber
\end{eqnarray}}\pagebreak[1]With
 the standard definition of the polarization basis tensors,
\begin{displaymath}
e_{+} \equiv \widehat{e}_\theta\otimes\widehat{e}_\theta -
\widehat{e}_\varphi\otimes\widehat{e}_\varphi,
\end{displaymath}
\begin{displaymath}
e_{\times} \equiv\widehat{e}_\theta\otimes\widehat{e}_\varphi -
\widehat{e}_\varphi\otimes\widehat{e}_\theta,
\end{displaymath}
the gravitational radiation waveform can then be written
\begin{equation}
h^{TT} = h_+e_{+} + h_\times e_{\times},
\label{htt}
\end{equation}
where the wave amplitudes are given by
\begin{eqnarray}
h_+&=&\frac{1}{r}\frac{G}{c^4}( {\skew6\ddot{
{I\mkern-6.8mu\raise0.3ex\hbox{-}}}}_{\theta\theta}- {\skew6\ddot{
{I\mkern-6.8mu\raise0.3ex\hbox{-}}}}_{\phi\phi}),
\label{hplus}\\
h_\times &=&\frac{2}{r}\frac{G}{c^4} {\skew6\ddot{
{I\mkern-6.8mu\raise0.3ex\hbox{-}}}}_{\theta\phi}.\label{hcross}
\end{eqnarray}

All of these physical quantities associated with gravitational
radiation are functions of at least the second time derivative of the
reduced quadrupole tensor ${I\mkern-6.8mu\raise0.3ex\hbox{-}}_{ab}$.
The standard quadrupole formula (SQF) consists of using these
relations in conjunction with the definition (\ref{ibar}) for
${I\mkern-6.8mu\raise0.3ex\hbox{-}}_{ab}$.  In an Eulerian code
${I\mkern-6.8mu\raise0.3ex\hbox{-}}_{ab}$ may be calculated directly
by summing over the grid, and the time derivatives calculated by using
the finite difference approximation $d/dt \rightarrow \Delta/\Delta
t$.  However, the successive application of numerical time derivatives
can introduce a great deal of noise into the calculated quantities,
especially when the time step varies from cycle to cycle.  To reduce
this problem, Finn \& Evans (1990) have developed two partially
integrated versions of the SQF that eliminate one of the time
derivatives by calculating
$\skew6\dot{I\mkern-6.8mu\raise0.3ex\hbox{-}}_{ab}$ directly.  They
find that both of these methods produce much cleaner waveforms, with
the high frequency numerical noise greatly suppressed.

If we take the time derivative of (\ref{ibar}) and use the mass
continuity equation~(\ref{mass-cont-L}) we obtain, after some
manipulation,
\begin{equation}
\skew6\dot{I\mkern-6.8mu\raise0.3ex\hbox{-}}_{ab} = -\int
\left(x_ax_b-\textstyle{\frac{1}{3}}
r^2\delta_{ab}\right)\nabla\cdot(\rho\vec{v})dV.
\label{qf2}
\end{equation}
This is the momentum divergence formula, which Finn \& Evans (1990)
refer to as QF2.  Analytically, using (\ref{qf2}) to calculate
gravitational waveforms and luminosities is equivalent to using
(\ref{ibar}) directly; however, in a finite difference formulation the
elimination of one numerical time derivative greatly increases the
signal-to-noise ratio.  We note that~(\ref{qf2}) introduces a spatial
derivative, the momentum divergence, into the calculation, and that
this derivative must be performed numerically.  While this can also be
expected to produce numerical noise, it should have little effect on
the resulting value for
$\skew6\dot{I\mkern-6.8mu\raise0.3ex\hbox{-}}_{ab}$ since the spatial
derivative appears inside the volume integral and the noise will be
smoothed by the integration process.

The divergence in (\ref{qf2}) can be eliminated by integrating by
parts and discarding the surface terms at infinity.  The result is
\begin{equation}
\skew6\dot{I\mkern-6.8mu\raise0.3ex\hbox{-}}_{ab} = 2\int \rho
\left[v_{(a}x_{b)}-\textstyle{\frac{1}{3}}
\delta_{ab}(\vec{v}\cdot\vec{x})\right]dV,
\label{qf1}
\end{equation}
where
\begin{displaymath}
v_{(a}x_{b)}=\frac{1}{2}\left(v_ax_b+v_bx_a\right) .
\end{displaymath}
Finn \& Evans (1990) refer to this as the first moment of momentum
formula, QF1.  Again, the use of (\ref{qf1}) in calculating the
gravitational radiation is mathematically equivalent to using
(\ref{ibar}) or (\ref{qf2}), but in a finite difference code QF1 has
several advantages.  The first is that, like QF2, one numerical time
derivative is eliminated when gravitational waveforms are calculated
using QF1.  The numerical divergence that must be calculated in QF2
has also been eliminated.  Another advantage is that the moment arm
that weights the mass has been reduced from $r^2$ to $r$, which
decreases the emphasis on low density material at large radii, while
the presence of the velocity increases the emphasis on the more
dynamical regions (Finn \& Evans 1990).

\subsection{Numerical Implementation of Quadrupole Formulae}
\label{sec-ngw}

We have implemented SQF, QF1, and QF2 in our hydrodynamics code to
calculate the gravitational radiation.  We calculate
${I\mkern-6.8mu\raise0.3ex\hbox{-}}_{ab}$ (for SQF) and
$\skew6\dot{I\mkern-6.8mu\raise0.3ex\hbox{-}}_{ab}$ (for QF1 and QF2)
on each of the cycles during the run and write them to disk for
additional post-processing to determine waveforms and luminosities.
Note that on cycle $n$ SQF produces quadrupole tensor components
defined at time level $t^{n+1}$, while QF1 and QF2 produce time
derivatives defined at level $t^{n+\frac{1}{2}}$.

The numerical integration of the SQF (\ref{ibar}) is a simple task.
We begin by defining\pagebreak[1] {\samepage\begin{eqnarray}
r^2_{j+\frac{1}{2},k+\frac{1}{2}}&=&\varpi^2_{j+\frac{1}{2}} +
z^2_{k+\frac{1}{2}},
\nonumber\\
x_{j+\frac{1}{2},l+\frac{1}{2}}&=&\varpi_{j+\frac{1}{2}}\cos\varphi_{l+\frac{1}{2}},
\label{xdefs}\\
y_{j+\frac{1}{2},l+\frac{1}{2}}&=&\varpi_{j+\frac{1}{2}}\sin\varphi_{l+\frac{1}{2}}.
\nonumber
\end{eqnarray}}\pagebreak[1]
The Cartesian components of ${I\mkern-6.8mu\raise0.3ex\hbox{-}}_{ab}$
are then
\begin{equation}
{I\mkern-6.8mu\raise0.3ex\hbox{-}}_{xx} = 2\sum_{j,k,l}
\rho_{{j+\frac{1}{2},k+\frac{1}{2},l+\frac{1}{2}}}\left(x^2_{j+\frac{1}{2},l+\frac{1}{2}}
-\textstyle{\frac{1}{3}}
r^2_{j+\frac{1}{2},k+\frac{1}{2}}\right)\Delta
V_{{j+\frac{1}{2},k+\frac{1}{2},l+\frac{1}{2}}},
\label{ixx}
\end{equation}
\begin{equation}
{I\mkern-6.8mu\raise0.3ex\hbox{-}}_{yy} = 2\sum_{j,k,l}
\rho_{{j+\frac{1}{2},k+\frac{1}{2},l+\frac{1}{2}}}\left(y^2_{j+\frac{1}{2},l+\frac{1}{2}}
-\textstyle{\frac{1}{3}}
r^2_{j+\frac{1}{2},k+\frac{1}{2}}\right)\Delta
V_{{j+\frac{1}{2},k+\frac{1}{2},l+\frac{1}{2}}},
\label{iyy}
\end{equation}
\begin{equation}
{I\mkern-6.8mu\raise0.3ex\hbox{-}}_{zz} = 2\sum_{j,k,l}
\rho_{{j+\frac{1}{2},k+\frac{1}{2},l+\frac{1}{2}}}\left(z^2_{k+\frac{1}{2}}
-\textstyle{\frac{1}{3}}
r^2_{j+\frac{1}{2},k+\frac{1}{2}}\right)\Delta
V_{{j+\frac{1}{2},k+\frac{1}{2},l+\frac{1}{2}}},
\end{equation}
\begin{equation}
{I\mkern-6.8mu\raise0.3ex\hbox{-}}_{xy} =
{I\mkern-6.8mu\raise0.3ex\hbox{-}}_{yx} = 2\sum_{j,k,l}
\rho_{{j+\frac{1}{2},k+\frac{1}{2},l+\frac{1}{2}}}
(xy)_{j+\frac{1}{2},l+\frac{1}{2}}
\Delta V_{{j+\frac{1}{2},k+\frac{1}{2},l+\frac{1}{2}}} ,
\end{equation}
and
\begin{equation}
{I\mkern-6.8mu\raise0.3ex\hbox{-}}_{xz}=
{I\mkern-6.8mu\raise0.3ex\hbox{-}}_{zx}=
{I\mkern-6.8mu\raise0.3ex\hbox{-}}_{yz}=
{I\mkern-6.8mu\raise0.3ex\hbox{-}}_{zy}=0,
\label{ixz}
\end{equation}
where we have used the fact that the summation is over the top half of
the distribution only ($z\geq0$) with reflection symmetry through the
$x-y$ plane.  The volume element is defined by (\ref{dvol}).  The
reduced quadrupole tensor calculated from (\ref{ixx})--(\ref{ixz}) is
mathematically traceless, as required.  In practice, however,
truncation errors can lead to a small non-zero trace if all components
are calculated as specified.  To guarantee the tracelessness of
${I\mkern-6.8mu\raise0.3ex\hbox{-}}_{ab}$, we replace the calculation
of ${I\mkern-6.8mu\raise0.3ex\hbox{-}}_{yy}$ using~(\ref{iyy}) with
\begin{equation}
{I\mkern-6.8mu\raise0.3ex\hbox{-}}_{yy} = -\left(
{I\mkern-6.8mu\raise0.3ex\hbox{-}}_{xx} +
{I\mkern-6.8mu\raise0.3ex\hbox{-}}_{zz} \right).
\end{equation}

The numerical implementation of QF2 presents some complications
because of the presence of both $\rho$ and $\vec{v}$, which are
defined in different positions on the grid, in the divergence.  Recall
from
\S~\ref{sec-grid-cen} above that the density is zone centered, while
each component of the velocity is face centered {\it on a different
face}.  In addition, zone centered quantities and face centered
quantities are defined at different time levels.  In order to obtain
an accurate discrete representation of (\ref{qf2}) we must perform the
summation on objects that are defined at the same space-time points.

At the time that we calculate the gravitational radiation, the density
has been updated to time level $n+1$ and the velocities have been
updated to time level $n+\frac{1}{2}$.  By saving the value of the
density at the previous time step we can interpolate the density to
the same time level as the velocity,
\begin{equation}
\rho^{n+\frac{1}{2}}_{{j+\frac{1}{2},k+\frac{1}{2},l+\frac{1}{2}}} =
\textstyle{\frac{1}{2}}\left(\rho^n_{{j+\frac{1}{2},k+\frac{1}{2},l+\frac{1}{2}}}+
\rho^{n+1}_{{j+\frac{1}{2},k+\frac{1}{2},l+\frac{1}{2}}}
\right).
\label{rhohalf}
\end{equation}
Here we use the simple average because time level $n+\frac{1}{2}$ is
defined to be halfway between levels $n$ and $n+1$,
\begin{displaymath}
t^{n+\frac{1}{2}}=\textstyle{\frac{1}{2}}\left(t^n+t^{n+1}\right).
\end{displaymath}
With this, all of the variables needed for QF2 are available at the
same time level.

To evaluate the divergence we use the identity
$\nabla\cdot(\rho\vec{v}) = \rho\nabla\cdot\vec{v} +
\vec{v}\cdot\nabla\rho$.  The divergence of $\vec{v}$ is naturally
defined at the zone center,
\pagebreak[1]{\samepage\begin{eqnarray}
\left(\nabla\cdot\vec{v}\right)_{{j+\frac{1}{2},k+\frac{1}{2},l+\frac{1}{2}}}
&=&
\frac{\varpi_{j+1}v_{j+1,k+\frac{1}{2},l+\frac{1}{2}}-\varpi_{j}v_{j,k+\frac{1}{2},l+\frac{1}{2}}}
{\varpi_{j+\frac{1}{2}}\Delta \varpi_{j+\frac{1}{2}}}
\nonumber\\&&\;\;\;\;\; +
\frac{u_{j+\frac{1}{2},k+1,l+\frac{1}{2}}-u_{j+\frac{1}{2},k,l+\frac{1}{2}}}{\Delta z_{k+\frac{1}{2}}}
\nonumber\\&&\;\;\;\;\;\;\;\;\; +
\frac{w_{j+\frac{1}{2},k+\frac{1}{2},l+1}-w_{j+\frac{1}{2},k+\frac{1}{2},l}}
{\varpi\Delta\varphi_{l+\frac{1}{2}}},\label{div-v}
\end{eqnarray}}\pagebreak[1]
while each component of the gradient of $\rho$ is naturally face
centered in that coordinate,
\begin{equation}
(\nabla_\varpi\rho)_{j,k+\frac{1}{2},l+\frac{1}{2}} =
\frac{\rho_{{j+\frac{1}{2},k+\frac{1}{2},l+\frac{1}{2}}}-\rho_{j-\frac{1}{2},k+\frac{1}{2},l+\frac{1}{2}}}{\Delta\varpi_{j}},
\end{equation}
\begin{equation}
(\nabla_z\rho)_{j+\frac{1}{2},k,l+\frac{1}{2}} =
\frac{\rho_{{j+\frac{1}{2},k+\frac{1}{2},l+\frac{1}{2}}}-\rho_{j+\frac{1}{2},k-\frac{1}{2},l+\frac{1}{2}}}{\Delta z_{k}},
\end{equation}
and
\begin{equation}
(\nabla_\varphi\rho)_{j+\frac{1}{2},k+\frac{1}{2},l} =
\frac{\rho_{{j+\frac{1}{2},k+\frac{1}{2},l+\frac{1}{2}}}-\rho_{j+\frac{1}{2},k+\frac{1}{2},l-\frac{1}{2}}}{\varpi_j\Delta\varphi_{l}}.
\label{grad-t-d}
\end{equation}
The integral in (\ref{qf2}) thus naturally breaks into four separate
discrete summations over the numerical grid,\pagebreak[1]
{\samepage\begin{eqnarray}
\skew6\dot{I\mkern-6.8mu\raise0.3ex\hbox{-}}_{ab}&=&
2\sum_{j,k,l} \left[(x_ax_b-\textstyle{\frac{1}{3}} r^2\delta_{ab})
\rho(\nabla\cdot\vec{v})\Delta
V\right]_{{j+\frac{1}{2},k+\frac{1}{2},l+\frac{1}{2}}}
\nonumber\\&& \;\;+\;\;
2\sum_{j,k,l} \left[(x_ax_b-\textstyle{\frac{1}{3}} r^2\delta_{ab})
v(\nabla_\varpi\rho)\Delta V\right]_{j,k+\frac{1}{2},l+\frac{1}{2}}
\nonumber\\&& \;\;+\;\;
2\sum_{j,k,l} \left[(x_ax_b-\textstyle{\frac{1}{3}} r^2\delta_{ab})
u(\nabla_z\rho)\Delta V\right]_{j+\frac{1}{2},k,l+\frac{1}{2}}
\nonumber\\&& \;\;+\;\;
2\sum_{j,k,l} \left[(x_ax_b-\textstyle{\frac{1}{3}} r^2\delta_{ab})
w(\nabla_\varphi\rho)\Delta V\right]_{j+\frac{1}{2},k+\frac{1}{2},l},
\label{nqf2}
\end{eqnarray}}\pagebreak[1]
where the factor of two appears to account for integration below the $x-y$
plane.  We again use the symmetry of the grid to write
\begin{equation}
\skew6\dot{I\mkern-6.8mu\raise0.3ex\hbox{-}}_{xz}=
\skew6\dot{I\mkern-6.8mu\raise0.3ex\hbox{-}}_{zx}=
\skew6\dot{I\mkern-6.8mu\raise0.3ex\hbox{-}}_{yz}=
\skew6\dot{I\mkern-6.8mu\raise0.3ex\hbox{-}}_{zy}=0
\label{ref-sym}
\end{equation}
and the tracelessness of the quadrupole tensor to obtain
\begin{equation}
\skew6\dot{I\mkern-6.8mu\raise0.3ex\hbox{-}}_{yy} = -\left(
\skew6\dot{I\mkern-6.8mu\raise0.3ex\hbox{-}}_{xx} + \
\skew6\dot{I\mkern-6.8mu\raise0.3ex\hbox{-}}_{zz} \right).
\label{traceless}
\end{equation}
We therefore use (\ref{nqf2}) to obtain only the three components
$\skew6\dot{I\mkern-6.8mu\raise0.3ex\hbox{-}}_{xx}$,
$\skew6\dot{I\mkern-6.8mu\raise0.3ex\hbox{-}}_{zz}$, and
$\skew6\dot{I\mkern-6.8mu\raise0.3ex\hbox{-}}_{xy}=
\skew6\dot{I\mkern-6.8mu\raise0.3ex\hbox{-}}_{yx}$.  Explicitly,
these are\pagebreak[1] {\samepage\begin{eqnarray}
\skew6\dot{I\mkern-6.8mu\raise0.3ex\hbox{-}}_{xx}&=&
2\sum_{j,k,l}
\left\{(x^2_{j+\frac{1}{2},l+\frac{1}{2}}-\textstyle{\frac{1}{3}}
r^2_{j+\frac{1}{2},k+\frac{1}{2}})
\left[\rho(\nabla\cdot\vec{v})\Delta
V\right]_{{j+\frac{1}{2},k+\frac{1}{2},l+\frac{1}{2}}}\right\}
\nonumber\\&& \;+\;
2\sum_{j,k,l} \left\{(x^2_{j,l+\frac{1}{2}}-\textstyle{\frac{1}{3}}
r^2_{j,k+\frac{1}{2}})
\left[v(\nabla_\varpi\rho)\Delta
V\right]_{j,k+\frac{1}{2},l+\frac{1}{2}}\right\}
\nonumber\\&& \;+\;
2\sum_{j,k,l}
\left\{(x^2_{j+\frac{1}{2},l+\frac{1}{2}}-\textstyle{\frac{1}{3}}
r^2_{j+\frac{1}{2},k})
\left[u(\nabla_z\rho)\Delta V\right]_{j+\frac{1}{2},k,l+\frac{1}{2}}\right\}
\nonumber\\&& \;+\;
2\sum_{j,k,l} \left\{(x^2_{j+\frac{1}{2},l}-\textstyle{\frac{1}{3}}
r^2_{j+\frac{1}{2},k+\frac{1}{2}})
\left[w(\nabla_\varphi\rho)\Delta
V\right]_{j+\frac{1}{2},k+\frac{1}{2},l}\right\},
\end{eqnarray}}\pagebreak[1]
{\samepage\begin{eqnarray}
\skew6\dot{I\mkern-6.8mu\raise0.3ex\hbox{-}}_{zz}&=&
2\sum_{j,k,l} \left\{(z^2_{k+\frac{1}{2}}-\textstyle{\frac{1}{3}}
r^2_{j+\frac{1}{2},k+\frac{1}{2}})
\left[\rho(\nabla\cdot\vec{v})\Delta
V\right]_{{j+\frac{1}{2},k+\frac{1}{2},l+\frac{1}{2}}}\right\}
\nonumber\\&& \;+\;
2\sum_{j,k,l} \left\{(z^2_{k+\frac{1}{2}}-\textstyle{\frac{1}{3}}
r^2_{j,k+\frac{1}{2}})
\left[v(\nabla_\varpi\rho)\Delta
V\right]_{j,k+\frac{1}{2},l+\frac{1}{2}}\right\}
\nonumber\\&& \;+\;
2\sum_{j,k,l} \left\{(z^2_{k}-\textstyle{\frac{1}{3}}
r^2_{j+\frac{1}{2},k})
\left[u(\nabla_z\rho)\Delta V\right]_{j+\frac{1}{2},k,l+\frac{1}{2}}\right\}
\nonumber\\&& \;+\;
2\sum_{j,k,l} \left\{(z^2_{k+\frac{1}{2}}-\textstyle{\frac{1}{3}}
r^2_{j+\frac{1}{2},k+\frac{1}{2}})
\left[w(\nabla_\varphi\rho)\Delta
V\right]_{j+\frac{1}{2},k+\frac{1}{2},l}\right\},
\end{eqnarray}}\pagebreak[1]
and\pagebreak[1] {\samepage\begin{eqnarray}
\skew6\dot{I\mkern-6.8mu\raise0.3ex\hbox{-}}_{xy}&=&
2\sum_{j,k,l} \left\{(xy)_{j+\frac{1}{2},l+\frac{1}{2}}
\left[\rho(\nabla\cdot\vec{v})\Delta
V\right]_{{j+\frac{1}{2},k+\frac{1}{2},l+\frac{1}{2}}}\right\}
\nonumber\\&& \;+\;
2\sum_{j,k,l} \left\{(xy)_{j,l+\frac{1}{2}}
\left[v(\nabla_\varpi\rho)\Delta
V\right]_{j,k+\frac{1}{2},l+\frac{1}{2}}\right\}
\nonumber\\&& \;+\;
2\sum_{j,k,l} \left\{(xy)_{j+\frac{1}{2},l+\frac{1}{2}}
\left[u(\nabla_z\rho)\Delta V\right]_{j+\frac{1}{2},k,l+\frac{1}{2}}\right\}
\nonumber\\&& \;+\;
2\sum_{j,k,l} \left\{(xy)_{j+\frac{1}{2},l}
\left[w(\nabla_\varphi\rho)\Delta
V\right]_{j+\frac{1}{2},k+\frac{1}{2},l}\right\}.
\end{eqnarray}}\pagebreak[1]
Here the divergence and gradient terms are obtained using
(\ref{div-v})--(\ref{grad-t-d}). The volume element, $\Delta V$, and
the position variables, $x$, $y$, and $r$, are defined by (\ref{dvol})
and (\ref{xdefs}), respectively, with appropriate indices.

The numerical representation of QF1 is also complicated by the
presence of both zone centered and face centered quantities in the
integral.  In this case, however, it is not possible simply to break
up the summation into zone centered and face centered parts.  It is
therefore necessary to interpolate one of the quantities onto the
position of the other.  We first interpolate the density to the same
time level as the velocities using (\ref{rhohalf}).  We then have the
choice of evaluating the density on the zone faces, or interpolating
the velocity components to the zone center.

As with QF2 the symmetries of the grid and the tracelessness of
${I\mkern-6.8mu\raise0.3ex\hbox{-}}_{ab}$ guarantee that
(\ref{ref-sym}) and (\ref{traceless}) still hold.  The Cartesian
velocity components in the $x$ and $y$-directions are\pagebreak[1]
{\samepage\begin{eqnarray} v_x&=&v\cos\varphi - w\sin\varphi,
\nonumber\\
v_y&=&v\sin\varphi + w\cos\varphi.
\nonumber
\end{eqnarray}}\pagebreak[1]
The zone centered representation of the components of
$\skew6\dot{I\mkern-6.8mu\raise0.3ex\hbox{-}}_{ab}$ is thus
\begin{equation}
\skew6\dot{I\mkern-6.8mu\raise0.3ex\hbox{-}}_{xx}=4\sum_{j,k,l}\left\{\rho
\left[\varpi\cos\varphi(v\cos\varphi-w\sin\varphi)-\textstyle{\frac{1}{3}}(v\varpi+uz)\right]
\Delta V\right\}_{{j+\frac{1}{2},k+\frac{1}{2},l+\frac{1}{2}}},
\end{equation}
\begin{equation}
\skew6\dot{I\mkern-6.8mu\raise0.3ex\hbox{-}}_{xy}=4\sum_{j,k,l}\left\{\rho
\left[\varpi\cos\varphi(v\sin\varphi+\textstyle{\frac{1}{2}} w\cos\varphi)
-\textstyle{\frac{1}{2}}\varpi w\sin^2\varphi\right]
\Delta V\right\}_{{j+\frac{1}{2},k+\frac{1}{2},l+\frac{1}{2}}},
\end{equation}
and
\begin{equation}
\skew6\dot{I\mkern-6.8mu\raise0.3ex\hbox{-}}_{zz}=4\sum_{j,k,l}\left\{\rho
\left[uz-\textstyle{\frac{1}{3}}(v\varpi+uz)\right]
\Delta V\right\}_{{j+\frac{1}{2},k+\frac{1}{2},l+\frac{1}{2}}},
\end{equation}
where the zone centered velocity components are calculated by volume
averaging as described in \S~\ref{sec-grid-cen}.  A face centered
discrete representation of (\ref{qf1}) can also be obtained.  In this
case the summation again breaks up into terms centered on each face,
in a manner similar to the separation of terms in QF2.  The mass
density is evaluated on each zone face by volume averaging, and we
obtain\pagebreak[1] {\samepage\begin{eqnarray}
\skew6\dot{I\mkern-6.8mu\raise0.3ex\hbox{-}}_{xx}&=&
4\sum_{j,k,l} \left\{\rho
\left[\varpi v\cos^2\varphi -\textstyle{\frac{1}{3}}\varpi v\right]\Delta V
\right\}_{j,k+\frac{1}{2},l+\frac{1}{2}}
\nonumber\\&& \;-\;
4\sum_{j,k,l} \left\{\textstyle{\frac{1}{3}}\rho zu\Delta
V\right\}_{j+\frac{1}{2},k,l+\frac{1}{2}}
\nonumber\\&& \;-\;
4\sum_{j,k,l}
\left\{\rho\varpi w
\cos\varphi\sin\varphi\right\}_{j+\frac{1}{2},k+\frac{1}{2},l},
\end{eqnarray}}\pagebreak[1]
{\samepage\begin{eqnarray}
\skew6\dot{I\mkern-6.8mu\raise0.3ex\hbox{-}}_{xy}&=&
4\sum_{j,k,l} \left\{\rho\varpi v\cos\varphi\sin\varphi\Delta V
\right\}_{j,k+\frac{1}{2},l+\frac{1}{2}}
\nonumber\\&& \;+\;
4\sum_{j,k,l}\left\{\textstyle{\frac{1}{2}}
\rho\varpi
w(\cos^2\varphi-\sin^2\varphi)\right\}_{j+\frac{1}{2},k+\frac{1}{2},l},
\end{eqnarray}}\pagebreak[1]
and
\begin{equation}
\skew6\dot{I\mkern-6.8mu\raise0.3ex\hbox{-}}_{zz}=
-4\sum_{j,k,l} \left\{\textstyle{\frac{1}{3}}\rho\varpi v\Delta
V\right\}_{j,k+\frac{1}{2},l+\frac{1}{2}} +4\sum_{j,k,l}
\left\{{\textstyle\frac{2}{3}}
\rho zu\Delta V\right\}_{j+\frac{1}{2},k,l+\frac{1}{2}}.
\end{equation}
We have implemented both the zone centered and the face centered
methods, and found no noticeable difference in the results.

The Finn \& Evans (1990) formulae QF1 and QF2 produce much less noisy
waveforms than the SQF.  Nevertheless, when luminosities, which
require the third derivative of the reduced quadrupole tensor, are
calculated using these expressions we still find that numerical noise
frequently dominates the signal.  This is especially true when the
time step is changing significantly from cycle to cycle.  We have
therefore implemented some techniques to improve the signal-to-noise
ratio when the numerical derivatives are taken.

The simplest method of taking a numerical time derivative is a simple
two-point formula,
\begin{displaymath}
\skew6\dot{I\mkern-6.8mu\raise0.3ex\hbox{-}}_{ab}^{n+\frac{1}{2}} =
\frac{{I\mkern-6.8mu\raise0.3ex\hbox{-}}_{ab}^{n+1}-
{I\mkern-6.8mu\raise0.3ex\hbox{-}}_{ab}^{n}}{\Delta
t^{n+\frac{1}{2}}}.
\end{displaymath}
When $\Delta t$ changes a lot from cycle to cycle, this can fluctuate
wildly from one time level to the next, causing numerical noise in the
signal.  These fluctuations can be reduced somewhat by obtaining the
derivative from a higher order fit to the neighboring points.  We have
implemented a 4-point cubic fit, as follows.  First
define\pagebreak[1] {\samepage\begin{eqnarray}
\delta t^{mn}& \equiv &t^m-t^n,
\nonumber\\
\delta{I\mkern-6.8mu\raise0.3ex\hbox{-}}^{mn}_{ab}
&\equiv&{I\mkern-6.8mu\raise0.3ex\hbox{-}}^m_{ab}-
{I\mkern-6.8mu\raise0.3ex\hbox{-}}^n_{ab}.
\nonumber\end{eqnarray}}\pagebreak[1]The time derivative can then be written
\begin{equation}
\skew6\dot{I\mkern-6.8mu\raise0.3ex\hbox{-}}_{ab}^n = \frac{\delta
{I\mkern-6.8mu\raise0.3ex\hbox{-}}_{ab}^{mn}}{\delta t^{mn}} - A\delta
t^{mn} -B(\delta t_{mn})^2,
\label{cub-fit}
\end{equation}
where
\begin{equation}
A = \frac {\delta {I\mkern-6.8mu\raise0.3ex\hbox{-}}_{ab}^{mn}/\delta
t^{mn}-\delta {I\mkern-6.8mu\raise0.3ex\hbox{-}}_{ab}^{pn}/\delta
t^{pn}} {\delta t^{mn}-\delta t^{pn}} - B(\delta t^{mn}-\delta
t^{pn}),
\end{equation}
and
\begin{equation}
B = \frac{\frac {\delta
{I\mkern-6.8mu\raise0.3ex\hbox{-}}_{ab}^{mn}/\delta t^{mn}-\delta
{I\mkern-6.8mu\raise0.3ex\hbox{-}}_{ab}^{pn}/\delta t^{pn}} {\delta
t^{mn}-\delta t^{pn}} - \frac {\delta
{I\mkern-6.8mu\raise0.3ex\hbox{-}}_{ab}^{mn}/\delta t^{mn}-\delta
{I\mkern-6.8mu\raise0.3ex\hbox{-}}_{ab}^{qn}/\delta t^{qn}} {\delta
t^{mn}-\delta t^{qn}}}{\delta t^{pn}-\delta t^{qn}}.
\label{cub-fit-B}
\end{equation}
This gives the time derivative at time level $t^n$.  The choice of the
time levels $m$, $p$, and $q$ in (\ref{cub-fit})--(\ref{cub-fit-B}) is
essentially arbitrary, although one should ideally choose nearby
points.  To provide additional noise reduction, we perform the
calculation for the four contiguous 4-point sets
$\{n,m,p,q\}=\{n,n+1,n+2,n+3\}$, $\{n,n+1,n+2,n-1\}$,
$\{n,n+1,n-1,n-2\}$, and $\{n,n-1,n-2,n-3\}$, and average the results
to obtain a value for the derivative.  This procedure is repeated for
each successive time derivative.

While using the cubic fit to obtain the derivative does provide a
cleaner waveform, there is still a significant amount of numerical
noise when several time derivatives are taken in succession.  We have
therefore found it necessary to pass the data through a filter to
smooth it after each derivative is performed.  The numerical noise in
the time derivative has a very high frequency $\sim1/\Delta t$,
whereas the signal that we are seeking is expected at much lower
frequencies $\sim1/t_D$, where
\begin{equation}
t_D = \left ( \frac{R^3}{GM} \right )^{1/2},
\label{tD}
\end{equation}
is the dynamical time, $R$ is the radius, and $M$ is the mass of the
system.  The relation $t_D > \Delta t$ is guaranteed because the
Courant conditions derived in \S~\ref{sec-courant} require that no
significant movement of the system can occur in a single time step.
We can therefore smooth the data using a low pass filter with a cutoff
frequency on the order of $\sim 1/\Delta t$, with little degradation
of the signal.  By doing this, we have been able to obtain smooth
functions for the waveforms and luminosities; see
\S~\ref{sec-ellip-tests} below.

There are many methods for filtering data (see, eg., Press et al.
(1992)).  Because we are usually dealing with a large number of
nonuniformly spaced data points, we use a ``real-time'' smoothing
technique that consists of performing a weighted average over a moving
window.  We have made it an option to choose either a recursive (input
includes filter output from previous cycles) or nonrecursive (input
includes only unfiltered data) filter, although we usually use the
recursive method.  We commonly use a triangular weighting function,
but other forms are available.  The size of the window is variable,
and is chosen to be at least several time steps, but smaller than any
expected signal features.  An estimate of the desired window size can
easily be obtained by looking at
${I\mkern-6.8mu\raise0.3ex\hbox{-}}_{ab}$, which has only small
amplitude numerical noise.

\section{CODE TESTS}\label{tests}

We have developed a 3-D code to be used as a computational laboratory
for modeling sources of gravitational radiation.  It has been
well-tested at all stages of its development to insure its accuracy
and verify its performance (Centrella, et al. 1986).  In this section
we discuss the major test-bed problems we used and present the results
of running them on our code.

\subsection{Riemann Shock Tube}\label{shock-tube}

The Riemann shock tube is a classic problem that tests the
hydrodynamics portion of the code, in particular the shock jump
conditions, shock heating, and the resolution of the shock and contact
discontinuities (Hawley, Smarr, \& Wilson 1984a,b).  Initially,
the system consists of a fluid at rest that exists in 2 states
separated by a membrane.  On the left of the membrane is hot, dense
fluid and on the right is cold, rarefied fluid.  At $t=0$ the membrane
is removed and the high density fluid flows into the low density
region, creating a shock front moving to the right and a rarefaction
wave moving to the left.

We ran the Riemann shock tube along all 3 coordinate directions. (For
the $\varpi$-direction we had to use a very large radius to
approximate 1-D planar symmetry.)  We tested several different
advection schemes and found that the LeBlanc monotonic method
discussed in \S~\ref{sec-mono} produced the best results; for this
reason we use it exclusively.  Figure~\ref{fig-shock-tube} shows
typical results from running a shock tube with 100 uniformly spaced
zones in the $\varpi$-direction.  The initial conditions for this run
are (Sod 1978):
\begin{eqnarray}
{\rm hot,\;dense\;gas:}& \;\;\;\; & \rho = 1.0 \; {\rm g/cm}^3
\nonumber \\ & \;\;\;\; & \varepsilon = 2.5 \; {\rm erg/g} \nonumber
\\ & \;\;\;\; & P = 1.0 \; {\rm erg/cm}^3 \nonumber \\ {\rm
cold,\;rarefied\;gas:}& \;\;\;\; & \rho = 0.125 \; {\rm g/cm}^3
\nonumber \\ & \;\;\;\; & \varepsilon = 2.0 \; {\rm erg/g} \nonumber
\\ & \;\;\;\; & P = 0.1 \; {\rm erg/cm}^3 , \nonumber
\end{eqnarray}
We used the perfect fluid EOS~(\ref{ideal-gas}) with $\gamma = 1.4$
and ran the shock tube in the $z$-direction.  As
Figure~\ref{fig-shock-tube} shows, the code results (shown by the
$\times$ symbols) accurately reproduce the analytic solution (shown by
the solid line).  Our results are essentially the same as those
obtained by Hawley, Smarr, \& Wilson (1984b) for monotonic advection.

\subsection{Spherical Polytropes}\label{sec-poly-tests}

A polytrope is a self-gravitating fluid that obeys the simple EOS
\begin{equation}
P=k\rho^\gamma=k\rho^{1+\frac{1}{n}},
\label{poly-eos}
\end{equation}
where $k$ is a constant related to the specific entropy and $n$ is
called the polytropic index.  This can be put into the form of an
ideal gas EOS (\ref{ideal-gas}) by defining the specific internal
energy to be
\begin{displaymath}
\varepsilon = \frac{k}{\gamma -1}\rho^{\gamma-1},
\end{displaymath}
where the $\gamma$'s that appear in (\ref{poly-eos}) and
(\ref{ideal-gas}) are chosen to be identical.  The smaller values of
$n$ produce ``stiffer'' equations of state, with $n=0$ corresponding
to an incompressible fluid.  A polytrope with index $n=1.5$
($\gamma=5/3$) corresponds to a classical ideal gas, while a polytrope
with $n=3$ ($\gamma=4/3$) models an ideal relativistic gas.  We have
used spherical polytropes to test the hydrodynamics coupled to
Newtonian gravity.  Since we have a cylindrical grid, this provides a
non-trivial test of our code.

\subsubsection{Stability and Symmetry}

The simplest test is to generate a stable spherical equilibrium
polytrope, and verify that the code preserves the initial
configuration.  We have run this test for a $\gamma=5/3$ polytrope on
a $100\times 100\times 64$ $(\varpi,z,\varphi)$ grid with zoning
ratios in the range 1.01 - 1.03.  We found that the models remained
stable for $> 10^6 t_D$.  We compared the gravitational potential
calculated by our Poisson solver with the analytic values and found
that the maximum deviation was $0.4\%$, just outside the surface of
the mass distribution, with errors $< 0.03\%$ throughout more than
$90\%$ of the zones.  The deviations from spherical symmetry were
$\sim 0.01\%$.

We found that an initially spherical polytrope placed on our numerical
grid exhibited small amplitude pulsations, indicating that the finite
difference model was not a true equilibrium configuration.  We
therefore inserted an artificial damping term into the velocity
equations and allowed the initial configuration to ``relax'' into
equilibrium.  The model that resulted when the oscillations had been
damped was slightly non-spherical.
The asphericity of the polytrope, in terms of the equatorial and
polar radii, is less
than a grid zone.  The moments of inertia about the $z$ and $x$-axes
differed by $0.44\%$ when the relaxation was performed on a $32\times
32$ ($\varpi,z$) 2-D axisymmetric grid.

\subsubsection{Homologous Collapse}

A polytrope with $n=3$ ($\gamma=4/3$) is marginally stable, and will
collapse homologously if the internal pressure is slightly reduced
(Goldreich \& Weber 1980). We have run such simulations in 2-D and
3-D, with moving and stationary grids of various sizes and zoning
ratios (Evans 1986).  In all cases, the collapse remained homologous
through increases in the central density of $10^4$--$10^5$, depending
on the resolution of the grid in the center of the polytrope.

Figure~\ref{poly.coll-2D} shows the results of a 2-D run with a
$100\times 100$ $(\varpi,z)$ grid and a zoning ratio of 1.02, and
Figure~\ref{poly.coll-3D} shows a 3-D run with a $40\times 40\times
20$ $(\varpi,z,\varphi)$ grid and a zoning ratio of 1.04; both models
use moving grids in the $\varpi$ and $z$ directions.  For these cases,
we set up a spherical polytrope with mass $8.4M_{\odot}$ and radius
$0.01R_{\odot}$.  Collapse was initiated by reducing pressure
throughout the polytrope by $1\%$ and imposing an inwardly directed
homologous velocity profile $v \propto r$ with maximum velocity
$10^{-10}$cm/sec.  The moving grid is homologous, and is specified by
choosing the grid velocity of a particular zone, $\sim 1/3$ of the way
from the center to the surface, to be a fixed fraction ($50\%$ for the
2-D run and $95\%$ for the 3-D run) of the fluid velocity in that
zone.

For homologous collapse the density profile, suitably scaled, should
not change throughout the evolution and the velocity should be purely
radial, $v \propto r$, and directed inward. Figure~\ref{poly.coll-2D}
shows $(a)$ the scaled density $\rho/\rho_c$ and $(b)$ the velocity
versus the scaled radial distance $(r/R_0)\rho_c^{1/3}$ for the 2-D
run.  Figure~\ref{poly.coll-3D} $(a)$ and $(b)$ show the same
quantities for the 3-D run. Here, $\rho_c$ is the central density and
$R_0$ is the initial polytrope radius. Each density graph includes
plots for $\log \rho_c$ = 9, 10, 11, 12, and 13; the 2-D plots in
Figure~\ref{poly.coll-2D} also show $\log \rho_c$ = 14.  We see that
the density profiles overlay each other within the resolution of the
graphics, except for the last trace on the graph for the 2-D collapse.
{}From the velocity plot for the 2-D collapse, Figure~\ref{poly.coll-2D}
$(b)$, we see that the core is beginning to rebound when
$\rho_c=10^{14}\hbox{g/cm}^3$, so we would not expect this density
profile to be the same as the other plots.  The velocity plots show a
homologous profile for the central region of the polytrope, including
more than $95\%$ of the mass, again except for the case
$\rho_c=10^{14}\hbox{g/cm}^3$ in the 2-D model.

The sphericity of the polytrope was well preserved throughout the
collapse in all runs.  Figure~\ref{poly-sphere} illustrates this for
the 3-D collapse described above.  Each graph displays three density
profiles, one for density along a cut along $\varpi$, one along the
$z$-axis, and one along the line $\varpi=z$.  We see that the density
profiles are indistinguishable from each other in both cases,
demonstrating that spherical symmetry has been well-preserved.

\subsection{Homogeneous Dust Ellipsoids}\label{sec-ellip-tests}

An initially uniform density, rigidly rotating, pressureless dust
ellipsoid will undergo gravitational collapse to a pancake while
maintaining a uniform density and an ellipsoidal figure.  This process
is described by a small number of ordinary differential equations
(Miller 1974; Saenz \& Shapiro 1978), which can be easily integrated
numerically to obtain a complete description of the collapse.  Since
the reduced quadrupole tensor for a uniform ellipsoid of axes $a$,
$b$, and $c$ aligned along the $x$, $y$, and $z$ axes, respectively,
is given by the simple expression
\begin{equation}
{I\mkern-6.8mu\raise0.3ex\hbox{-}}_{ab} =
\frac{M}{15}\mbox{diag}(2a^2-b^s-c^2,2b^2-a^s-c^2, 2c^2-a^s-b^2),
\end{equation}
the expected gravitational radiation from this collapse is easily
obtained using the quadrupole formula.  Finally, the gravitational
potential of a uniform density ellipsoid is also known theoretically
(Chandrasekhar 1969).  The collapse of such uniform ellipsoids thus
provides an important means of testing the hydrodynamics,
gravitational radiation, and Poisson solver of our code.

Figures~\ref{ellip-2D-1}-\ref{ellip-2D-2} illustrate the results of
simulations of the collapse of axisymmetric dust ellipsoids.  The
plots have
all been scaled to the natural scales of the problem.  Thus, the
axes $a=b$, and $c$ have been scaled by the initial length of
the semi-major axis $a_0$.  The times are expressed in units of
the dynamical time $t_D$ for a {\em spherical}
polytrope of the same mass $M$ and radius $R=a_0$; see
equation~(\ref{tD}).  Finally, the gravitational radiation
waveforms and efficiencies are scaled by $a_0/M^2$ and
$(M/a_0)^{-7/2}$, respectively.

Figure~\ref{ellip-2D-1} shows the behavior of an axisymmetric oblate
ellipsoid $a=b$ uniformly rotating about the $z$-axis with
$\Omega=0.634 t_D^{-1}$ and
eccentricity $e=0.1$.  This calculation was run using a 2-D $130 \times
130$ $(\varpi,z)$ grid with uniform zoning and moving grids.  The
evolution of the
semi-major (equatorial) and semi-minor (polar) axes is shown in
Figure~\ref{ellip-2D-1}~$(a)$, using solid lines for the theoretical
values and dashed lines for the code results.
Figure~\ref{ellip-2D-1}~$(b)$ shows the evolution of the eccentricity
throughout the collapse as the ellipsoid approaches a ``pancake''
with $e=1$.  Due to the axisymmetry,
${I\mkern-6.8mu\raise0.3ex\hbox{-}}_{xx}$ =
${I\mkern-6.8mu\raise0.3ex\hbox{-}}_{yy}$ = $- \frac{1}{2}
{I\mkern-6.8mu\raise0.3ex\hbox{-}}_{zz}$, and all the off-diagonal
components of ${I\mkern-6.8mu\raise0.3ex\hbox{-}}_{ab}$ are zero.
Thus, the gravitational radiation is all in the ``$+$'' polarization,
and the amplitude has a simple $\sin^2 \theta$ dependence.
The on-axis waveform $r\;h_+(\theta=\varphi=0)$ is shown in
Figure~\ref{ellip-2D-1} $(c)$.  Finally, Figure~\ref{ellip-2D-1} $(d)$
shows the energy
emitted as gravitational radiation $\Delta E/Mc^2$.  The gravitational
radiation waveforms and efficiency dispalyed here were
obtained using QF1a.  Those obtained using QF1b and QF2 are
indistinquishable, while the waveform obtained using SQF exhibits a
substantial amount of numerical noise.  Overall, we see
that the code matches the results of the theoretical calculation very
well.

In addition, we ran an axisymmetric collapse starting with
eccentricity $e=0.99$ and $\Omega=1.284 t_D^{-1}$ using a
$64 \times 32$ $(\varpi,z)$ grid with
uniform zoning and moving grids.  This eccentricity corresponds to an
axis ratio of $0.141$, significantly more oblate than the axis
ratio of $0.205$ used for the bar mode instability problem of the
next section.  The
results are shown in Figure~\ref{ellip-2D-2}, which gives $(a)$ the
behavior of the axes, $(b)$ the eccentricity, $(c)$ the waveform
$r\;h_+(\theta=\varphi=0)$,
and $(d)$ the energy emitted as gravitational radiation $\Delta
E/Mc^2$.  We see that the code results again agree closely with
the analytic results.  While the simulation results for the
hydrodynamic quantities, such as the axes, match the analytic results
well throughout the calculation, we can see that,
towards the end of the simulation, when the
$c$-axis is reduced to only a few zones in the $z$-direction, the
gravitational wave calculation breaks down due to
insufficient resolution.

\subsection{Bar Mode Instability}

Rotating axisymmetric fluids are subject to global rotational
instabilities that arise from growing modes $e^{im \phi}$, where $m=2$
is the so-called ``bar mode''.  These can be parametrized by
\begin{equation}
\beta = \frac{T}{|W|},
\label{T/W}
\end{equation}
where $T$ is the rotational kinetic energy and $W$ is the
gravitational potential energy; see Durisen \& Tohline (1985) and
Tassoul (1978) for reviews.  The dynamical bar instability is driven
by hydrodynamics with Newtonian gravity and develops on a timescale of
about a rotation period.  It occurs for values of $\beta >
\beta_d$, where $\beta_d \approx 0.27$ for a wide range of rotating
fluids (Durisen \& Tohline 1985; Shapiro \& Teukolsky 1983).  The
growth rate and eigenfrequency of the bar mode can be calculated using
the linearized tensor virial equations (TVE) (Chandrasekhar 1969).
The development of the bar instability in a differentially rotating
polytrope with $n = 3/2$ for various values of $\beta$ has been
studied extensively by Tohline, Durisen, \& McCollough (1985,
hereafter TDM) using 3-D numerical simulations which were compared to
the results of the TVE analysis.  Since both their numerical and
analytic results are available, we decided to use this case, with
$\beta = 0.30$, as a test bed problem for the hydrodynamics and
gravity.

We set up the initial axisymmetric equilibrium polytrope on a
cylindrical grid using the self-consistent field method of Smith \&
Centrella (1992), which is based on earlier work of Ostriker \& Mark
(1968) and Hachisu (1986).  We specify a rotation law of the general
form $j = j(m(\varpi))$, where $j$ is the specific angular momentum
and $m(\varpi)$ is the mass interior to a cylinder of radius $\varpi$
(Ostriker \& Mark 1968).  For comparison with previous work we use the
rotation law for the uniformly rotating, constant density Maclaurin
spheriods,
\begin{equation}
j(m) = {\textstyle{\frac{5}{2}}} (1 - (1 - m)^{2/3}) J/M,
\label{rotation-law}
\end{equation}
where $J$ is the total angular momentum and $M$ is the total mass.
Since polytropes do not have constant density, this produces
differential rotation.

The free parameters in this self-consistent field method are, in
addition to the rotation law $j(m)$, the polytropic index $n$, the
central density $\rho_c$, and the axis ratio $R_p/R$, where $R_p$ is
the polar radius and $R$ is the equatorial radius.  Upon convergence
to a solution of the equations of hydrodynamic equilibrium, this
method gives the total mass $M$, the polytropic constant $k$ (and
hence the pressure $P$), the angular velocity $\Omega(\varpi)$, and
$\beta$.  Since $\beta$ is not specified initially, some
experimentation with the input parameters is generally necessary to
achieve a particular value of $\beta$.  In constructing our model with
$\beta \approx 0.30$ we were guided in our choice of input parameters
by the values given in Tohline, Durisen, \& McCollough (1985).  We
found that using $R_p/R = 0.205$ gives $\beta = 0.301$.  Note that
this model is highly flattened due to the rotation.

The self-consistent field code produces a rotating axisymmetric
equilibrium using a relatively high resolution grid; this model is
then interpolated onto a lower resolution grid for evolution by the
3-D code.  To evolve the test run described here we used a $64 \times
32 \times 64$ $(\varpi,z,\varphi)$ grid.  Calculating $T$ and $W$ over
this grid at the initial time gives $\beta = 0.295$.  We set the grid
boundaries at $\varpi = 3.85 R$ and $z = 1.60R = 7.91 R_p$.  This
large amount of initially empty space is necessary both to provide
room for the star to expand as the bar mode grows as well as to the
specify the boundary conditions for the solution of Poisson's equation
accurately.  We used a finer grid in the region initially occupied by
the matter and a coarser grid outside.  The $\varpi$ grid is uniform
to the zone $j=30$, and the $z$ grid is uniform up to $k=16$.  The
zoning is chosen such that the center of the radial zone $j=25$ is at
$R$ and the center of the axial zone $k=9$ is at $R_p$.  Outside of
this uniformly zoned region, the zone size increases linearly with
zoning ratios $\Delta \varpi_{j+1}/ \Delta \varpi_j = 1.03$ and
$\Delta z_{k+1} / \Delta z_k = 1.1$ The angular grid is uniformly
zoned.  All grids are held fixed throughout this simulation.
To trigger the bar
instability we imposed a random perturbation with amplitude $10^{-3}$
on the density in each grid cell.

Figure~\ref{spiral-arms} shows the development of the bar mode at 4
different times during the evolution of this model.  The first
frame, at $18.94 t_D$, shows the model near the end of the linear
growth regime of the bar mode.  The bar structure is very apparent
here, and the model has not yet expanded significantly beyond
its initial boundaries.  At $20.02 t_D$ we see the beginning of the
growth of spiral arms.  By $23.26 t_D$ the model has expanded
significantly and the spiral arms are pronounced.  In the final
frame, at $26.76 t_D$, the core has returned to a more axisymmetric
configuration and the spiral arms have expanded to about twice the
initial radius.  When evolved further, the spiral arms merge
into a tenuous envelope surrounding the central core.
These results
are qualitatively similar to those obtained by TDM.  We observe
the formation of shocks during the formation and expansion of the
spiral arms, which were not modelled in the TDM simulation.

To study the development of the bar mode quantitataively, we analyze
the density in a ring of fixed $\varpi$ and $z$ using a complex
Fourier integral
\begin{equation}
C_m(\varpi,z) = \frac{1}{2 \pi}
\int_0 ^{2 \pi} \rho(\varpi,\varphi,z) e^{im\varphi} d\varphi,
\label{Cm}
\end{equation}
where $m = 2$.  The bar mode amplitude is then
\begin{equation}
|C| = |C_2|/C_0,
\label{bar-amp}
\end{equation}
where $C_0(\varpi,z) = \overline \rho(\varpi,z)$ is the mean density
in the ring, and the phase angle is
\begin{equation}
\phi = \tan^{-1}{\rm Re}(C_2)/{\rm Im}(C_2).
\label{phase}
\end{equation}
According to the TVE analysis,
the amplitude $|C|$ should grow exponentially with time as
the instability develops.

Figure~\ref{bar-mode} shows the growth of the bar mode $\ln |C|$
versus time for the ring $\varpi = 0.362 R_{eq}$ in zone $j = 10$
in the
equatorial plane $z=0$.  We find the slope $d\ln|C|/dt = 0.591
t_D^{-1}$.  This compares favorably with the TVE result $d\ln
|C|/dt = 0.728 \pm 0.038 t_D^{-1}$ given by TDM (we have converted
from their units).  For the eigenfrequency,
which is twice the pattern rotation speed we
find $\sigma_2 = 1.899 t_D^{-1}$, which is very close
to the TVE prediction
$\sigma_2 = 1.892 \pm 0.094 t_D^{-1}$ (TDM).

We can also compare our results with the numerical calculations of
TDM.  They used a 3-D Eulerian code written in cylindrical coordinates
and required that the EOS remain polytropic.  Their code does not
solve an energy equation, and therefore has no way to handle
self-consistently the shocks that form as the spiral arms expand. They
used donor cell advection, which is known to be very diffusive; see
\S~\ref{sec-mono}.  And, their code assumes ``$\pi$-symmetry'', which
means that the flow is calculated only in the angular range $0 \leq
\varphi < \pi$ so that only the even mode distortions are modeled.
They used a $31 \times 15 \times 32$ $(\varpi,z,\varphi)$ grid,
with the model extending out to zone 24 in the $\varpi$-direction
and zone 9 in the $z$-direction, which is equivalent in resolution
to our initial setup.
They found a bar mode growth rate of
$d\ln|C|/dt = 0.218$ and an eigenfrequency $\sigma_2 = 2.102$.  The
substantial deviation from the TVE result for the growth rate is
attributed to the large numerical diffusion in their code; see TDM for
details.  Overall, our code, which takes advantage of more modern
techniques, performs significantly better on this problem.

\section{SUMMARY}
\label{summary}

We have developed a 3-D Eulerian hydrodynamics code designed to model
sources of gravitational radiation.  This code is written in
cylindrical coordinates $(\varpi,z,\varphi)$, and has moving grids in
the $\varpi$ and $z$-directions. The hydrodynamical equations are
solved using operator splitting (Wilson 1979; Bowers \& Wilson 1991).
The equations are divided into a ``Lagrangian step'', which includes
Lagrangian-like processes such as accelerations and work done on a
fluid element, and a ``transport step'', which includes advection of
material through the grid and grid motion.  The transport is handled
using a monotonic advection scheme developed by LeBlanc (Clancy 1989;
Bowers \& Wilson 1991), with the consistent advection scheme for
angular momentum transport (Norman \& Winkler 1986; Norman, Wilson, \&
Barton 1980).

We use Newtonian gravity, and calculate the gravitational radiation
produced using the quadrupole approximation.  Poisson's equation is
solved using an efficient and robust 3-D matrix solver based on the
preconditioned conjugate gradient method.  We have implemented the
partially integrated quadrupole formulae QF1 and QF2 developed by Finn
\& Evans (1990) and numerical filtering techniques to produce smooth
gravitational waveforms and luminosities.

This code has been extensively tested to verify its accuracy and
stability.  For example, the Riemann shock tube has been used to test
the hydrodynamics part of the code, with particular attention to the
shock jump conditions, shock heating, and resolution of the shock and
contact discontinuities.  The homologous collapse of spherical
polytropes was used to test the Poisson solver and the coupling
between the hydrodynamics and gravity, as well as the ability of the
code to reproduce spherical symmetry on a cylindrical grid as the
polytrope collapsed through many order of magnitude in density.  The
gravitational collapse of axisymmetric homogeneous dust ellipsoids
expanded the tests to include the
gravitational radiation produced.  Finally, the development of the
dynamical bar mode instability was modeled, and the results compared
with theoretical calculations and previous numerical models.  In all,
our code has produced accurate and stable solution to these test-bed
problems, and we believe it is ready to be used to model more realistic
astrophysical sources of gravitational radiation.

\acknowledgments
We thank R. Bowers and J. LeBlanc for many enlightening discussions
and much valuable help.  This work was supported by NSF grants
PHY-8451732, PHY-8706135, PHY-9012383, and PHY-9208914, and by LANL.
The simulations were carried out at PSC, NCSA, and LANL.

\newpage

\begin{figure}[p]
\caption{Staggered z-grid illustrating
zone centered and face centered coordinate definitions.  Zone faces
are denoted by integral indices $z_k$ and zone centers by
half-integral indices $z_{k+\frac{1}{2}}$.
\label{z-gridfig}}
\end{figure}

\begin{figure}[p]
\caption{Centering of variables within a grid cell.  Scalar quantities
$S_{j+\frac{1}{2},k+\frac{1}{2},l+\frac{1}{2}}$ are defined in the
zone center.  The velocity components
$v_{\varpi}=v_{j,k+\frac{1}{2},l+\frac{1}{2}}$,
$v_{z}=u_{j+\frac{1}{2},k,l+\frac{1}{2}}$, and
$v_{\varphi}=w_{j+\frac{1}{2},k+\frac{1}{2},l}$ are defined on the
zone faces.  (Adapted from Tohline, Durisen \& McCollough (1985).)
\label{gridfig}}
\end{figure}

\begin{figure}[p]
\caption{Volume averaging of zone centered variable to achieve face
centered value.  Diagonally hatched regions indicate the area for
which $q_{j+\frac{1}{2},k+\frac{1}{2}}$ or
$q_{j+\frac{1}{2},k-\frac{1}{2}}$ is constant on the zone centered
grid.  Shaded region indicates the area on the face centered grid for
which $q_{j+\frac{1}{2},k}$ is to be evaluated.  The angular dimension
and index have been suppressed in the diagram for clarity.
\label{fig-average}}
\end{figure}

\begin{figure}[p]
\caption{Second order monotonic interpolation. The slopes $S_L$,
$S_R$, and $S_2$ are calculated as described in the text.  The
interpolated density $\rho^*_k$ is evaluated at a distance
$\frac{1}{2} u_k \Delta t$ ``upwind'' from the zone edge $z_k$.
\label{monofig}}
\end{figure}

\begin{figure}[p]
\caption{Riemann shock tube.  This test was run with 100 uniformly
spaced zones in the $\varpi$-direction using the initial conditions
given in the text.  The solid line shows the analytic solution, and
the $\times$ symbols show the code results.  Results are given at $t =
21.7$ sec. All quantities are plotted in cgs units.
\label{fig-shock-tube}}
\end{figure}

\begin{figure}[p]
\caption{Homologous collapse of polytrope in 2-D.  A
$100\times 100$ $(\varpi,z)$ moving grid was used.  Each graph
includes plots for $\log\rho_c=9,10,11,12,13$, and $14$.
\label{poly.coll-2D}}
\end{figure}

\begin{figure}[p]
\caption{Homologous collapse of polytrope in 3-D.  Here, a $40\times
40\times 20$ $(\varpi,z,\varphi)$ grid was used, with
moving grids in the $\varpi$ and $z$-directions.
Each graph includes plots for
$\log\rho_c=9,10,11,12$, and $13$.
\label{poly.coll-3D}}
\end{figure}

\begin{figure}[p]
\caption{Sphericity test for collapsing polytrope in 3-D.  In each
plot 3 density
profiles are shown, taken along a radial ($\varpi$) cut, along the
$z$-axis, and along the line $\varpi=z$.
\label{poly-sphere}}
\end{figure}

\begin{figure}[p]
\caption{Dynamics of a collapsing, uniform density, dust ellipsoid.
This figure shows an axisymmetric case with initial eccentricity
$e = 0.1$ rotating around the minor ($z$) axis with angular velocity
$\Omega=0.634 t_D^{-1}$,
run on a 2-D $130 \times 130$ $(\varpi,z)$~grid.  The solid
lines show the theoretical values, and the dashed lines show the
code results.  $(a)$~Evolution of the axes.  $(b)$~Eccentricity
$(c)$~Gravitational waveform.  $(d)$~Energy emitted as gravitational
radiation.
\label{ellip-2D-1}}
\end{figure}

\begin{figure}[p]
\caption{Same as Fig.~\protect{\ref{ellip-2D-1}}, but for an initial
eccentricity $e = 0.99$ and $\Omega=1.284 t_D^{-1}$,
run on a $64 \times 32$ $(\varpi,z)$ grid.
\label{ellip-2D-2}}
\end{figure}

\begin{figure}[p]
\caption{Development of the dynamical bar mode instability in a
differentially rotating polytrope with $\beta = 0.3$.  Density
contours are logarithmic and are plotted at density levels that are 1,
2, and 3 decades down from $\rho_c$.
$(a)$~$t = 18.94t_D$  $(b)$~$t = 20.02 t_D$
$(c)$~$t = 23.26 t_D$ $(d)$~$t = 26.76 t_D$ \label{spiral-arms}}
\end{figure}

\begin{figure}[p] \caption{The growth of the bar mode
instability.  The amplitude of the bar mode $\ln |C|$ at $\varpi =
0.362 R_{eq}$ is shown versus time.  The straight line has slope
$d\ln|C|/dt = 0.591 t_D^{-1}$.  \label{bar-mode}} \end{figure}
\end{document}